\documentstyle[aaspp4,12pt]{article}

\def\etal{{\it et al.$\,\,$}}
\def\simless{\mathbin{\lower 3pt\hbox
   {$\rlap{\raise 5pt\hbox{$\char'074$}}\mathchar"7218$}}} 
\def\simgreat{\mathbin{\lower 3pt\hbox
   {$\rlap{\raise 5pt\hbox{$\char'076$}}\mathchar"7218$}}} 

\def\rsun{{{R}_{\odot}}}
\def\vsun{{{\bf V}_{\odot}}}
\def\msun{{M_{\odot}}}
\def\eps02{{{\epsilon_{-2}}}}

\def\kms{{km s$^{-1}$ }}
\def\mathbf{{\boldmath}}

\def\mura{{\mu_{\rm RA} }}
\def\mudec{{\mu_{\rm DEC} }}
\def\errmura{{\sigma_{\mu_{RA}} }}
\def\errmudec{{\sigma_{\mu_{DEC}} }}

\def\muw{{{\mu}_W}}

\def\taus{{\tau_S}}
\def\vperp{{V_{\perp}}}
\def\vr{{V_r^{(P)}}}
\def\vri{{V_r}}
\def\vrd{{V_r}^{(DGR)}}

\def\vpmath{{ \mbox{\boldmath V}}_{\perp}^{(P)}}
\def\cl{{\cal L}}
\def\cd{{\cal D}}
\def\zb{{z_0}}

\def\dmuvec{{ \mbox{\boldmath $\delta\mu$} }}
\def\muvec{{ \mbox{\boldmath $\mu$} }}
\def\data{{\tilde \muvec}}
\def\dataw{{\tilde{\mu}_W}}

\def\errmu{{\sigma_{\muvec} }}

\def\thetavec{{ \mbox{\boldmath $\theta$} }}

\def\zhat{{\bf \hat z}}
\def\nhat{{\bf \hat n}}

\tighten
\singlespace
\begin{document}
\title{NEUTRON STAR POPULATION DYNAMICS.II: \\ 
3D SPACE VELOCITIES OF YOUNG PULSARS} 
\author{J. M. Cordes}
\affil{Astronomy Department and NAIC, Cornell University}
\author{David F. Chernoff}
\affil{Astronomy Department, Cornell University}
\begin{abstract}

We use astrometric, distance and spindown data on pulsars to: 
(1) estimate three-dimensional velocity
    components, birth distances from the galactic plane,
    and ages of individual objects; 
(2) determine the distribution of space velocities 
    and the scale height of pulsar progenitors; 
(3) test spindown laws for pulsars;
(4) test for correlations between space velocities and other 
    pulsar parameters; and
(5) place empirical requirements on mechanisms than can 
    produce high velocity neutron stars.
Our approach incorporates 
measurement errors, 
uncertainties in distances, 
deceleration in the Galactic potential, and 
differential galactic rotation.  
We focus on a sample of proper motion measurements of 
young pulsars ($<10$ Myr) whose
trajectories may be accurately and simply modeled.  This sample of 47
pulsars excludes millisecond pulsars and other objects that may have
undergone accretion-driven spinup. 

We estimate velocity components and birth $z$ distance on a
case-by-case basis assuming the actual age equals the conventional
spindown age for a breaking index n=3, no torque decay, and birth
periods much shorter than present day periods.  Every sample member
could have originated within 0.3 kpc of the galactic plane while still
having reasonable present day peculiar radial velocities.

For the 47-object sample, the scale height of the progenitors is $\sim
0.13$ kpc and the three dimensional velocities are distributed in two
components with characteristic speeds of $175^{+20}_{-30}$ km s$^{-1}$
and $700^{+200}_{-150}$ km s$^{-1}$, representing $\sim 83\%$ and
$\sim 17\%$ of the population, respectively.  The sample  
velocities are inconsistent with a single component Gaussian
model, well-described by a two component Gaussian model but do not
require models of additional complexity. From the best-fit
distribution, we estimate that about 20\% of the known pulsars will
escape the Galaxy, assuming an escape speed of 500 km s$^{-1}$. The
best-fit, dual-component model, if augmented by an additional,
low velocity ($<50$ km s$^{-1}$) component tolerates, at most, only
a small extra contribution in number, 
less than 5\%. The best three component models do not show a
preference for filling in the probability distribution at speeds
intermediate to 175 and 700 km s$^{-1}$ but are nearly degenerate with
the best two component models.   We estimate that the high velocity tail
($>1000$ km s$^{-1}$) may be underrepresented by a factor
$\sim 2.3$ owing to selection effects in pulsar surveys. 

The estimates of scale height and velocity parameters are insensitive
to the explicit relation of chronological and spindown ages.  A
further analysis starting from our inferred velocity distribution
allows us to test spindown laws and age estimates. There exist
comparably good descriptions of the data involving different
combinations of braking index and torque decay timescale. We find a
braking index of 2.5 is favored if torque decay occurs on a time scale
of $\sim 3$ Myr, while braking indices between 3.5 and 6 are preferred
if there is no torque decay.  For the sample as a whole, the
most-probable chronological ages are typically smaller than conventional
spindown ages by factors as large as two.

We have also searched for correlations between three-dimensional
speeds of individual pulsars and combinations of spin period and
period derivative.  None appears to be significant.  We argue that
correlations identified previously between velocity and (apparent)
magnetic moment reflect the different evolutionary paths
taken by young, isolated (nonbinary), high-field pulsars and older,
low-field pulsars that have undergone accretion-driven spinup.  We
conclude that any such correlation measures differences in spin and
velocity selection in the {\it evolution} of the two populations and
is not a measure of processes taking place in the core collapse that
produces neutron stars in the first place.  

We assess mechanisms for producing high-velocity neutron stars,
including disruption of binary systems by symmetric supernovae
and neutrino, baryonic or electromagnetic
rocket effects during or shortly after the supernova.  
The largest velocities seen ($\sim 1600$ \kms) along with the 
paucity of low-velocity pulsars suggest that disruption of
binaries by symmetric explosions is insufficient.   Rocket effects
appear to be a necessary and general phenomenon.
The required kick amplitudes and the absence
of a magnetic-field-velocity correlation do not yet rule out
any of the rocket models.  
However, the required amplitudes suggest that the core
collapse process in a supernova is highly dynamic and aspherical
and that the impulse delivered to the neutron star
is larger than existing simulations
of core collapse have achieved.

\end{abstract}


\section{INTRODUCTION}\label{sec:intro} 

The high velocity nature of pulsars has been recognized since the
first, minimal sample was analyzed by Gunn \& Ostriker (1970).  As
pulsar discoveries and proper-motion measurements have increased, and
the pulsar distance scale improved, the largest inferred velocity has
increased to $\simgreat 1000$ km s$^{-1}$.  The population's dispersion exceeds
that of any other normal Galactic stellar
population.

Although it is likely that a pulsar's characteristic motion is
generated during the neutron star formation, a number of physical
mechanisms has been proposed. In broad terms, these include: 
(1) disruption of binaries through instantaneous, symmetric mass loss in
supernova explosions (Blaauw 1961; Gott, Gunn \& Ostriker 1970;
Radhakrishnan \& Shukre 1985); 
(2) slow, post-explosion, rocket effects
associated with loss of spin energy by the neutron stars (e.g. 
Harrison \& Tademaru 1975; Helfand \& Tademaru 1977); and 
(3) an instantaneous momentum impulse, or ``kick'', imparted through
asymmetry of the supernova explosion (Shklovskii 1970;
Dewey \& Cordes 1987).

Many statistical analyses of pulsar velocities have attempted to
discern which (if any) of the mechanisms is favored (Dewey \& Cordes 1987;
Bailes 1989; Iben \& Tutukov 1996). The recent
treatment by Lyne \& Lorimer (1994; hereafter LL) has incorporated new
proper-motion measurements (Harrison \etal 1993) and an up-to-date
model for the free electron density in the Galaxy (Taylor \& Cordes
1993; hereafter TC). The latter is essential for estimating distances
to most pulsars.  LL find a larger mean pulsar speed,
$\overline{V} \sim $ 450 km s$^{-1}$, than had been estimated
previously. The probability density function (pdf) for the
perpendicular speed is
\begin{equation} 
f_{\vperp}(\vperp) \propto
\frac{(\vperp/V_0)^{0.3}}
                          {1+(\vperp/V_0)^{3.3}},
\end{equation}
where $V_0 = 330$ km s$^{-1}$ (and $dN \propto f_{\vperp} d^2\vperp$)
\footnote{The given expression corrects a typographical error in LL,
according to Mollerbach \& Roulet (1997)}.
LL state that $\sim 50$\% of pulsars will escape the Galaxy,
assuming an escape speed $\sim 500$ km s$^{-1}$.  LL also re-emphasized
the fact that pulsar surveys tend to undercount the highest velocity
pulsars. Surveys are signal-to-noise limited and sample
a finite volume which, of course, is roughly centered on the
Galactic plane.  Since it is generally thought that most pulsars are born in
the vicinity of the Galactic plane and move away from it,
fast objects spend less
time in the detectable volume than slow ones and are consequently
under-represented.

In attempting to characterize the pulsar velocity distribution, a
fundamental difficulty is that {\it all} of the basic components
necessary for a kinematic description of the population are poorly
known.  Except for a few historical supernovae, the {\it chronological age}
of each individual object is unknown and it must be estimated from the
``spindown'' age.  The {\it distance} is uncertain, though it may be
constrained by a pulsar's dispersion measure (DM) from knowledge of the
electron density within the Galaxy.  The perpendicular components of
the {\it velocity} may be derived from the proper motion and distance,
though the proper motion measurements only exist for a subset of the
entire pulsar population and both distance and proper motion
determinations have significant errors. In summary,
there are significant uncertainties in the age, distance and velocity
of nearly every known pulsar.

Our general goal is to infer properties, like the distribution of kick
velocities at birth, for the pulsar population. Since there is no {\it
a priori} way to separate the uncertainties we must strive to treat
them in an even-handed, consistent manner.  Although we do not
completely fulfill that goal in this paper, we make significant
progress by incorporating the uncertainties in position and velocity
directly into our analysis and by separately considering the
sensitivity of the results to the relation between spindown and
chronological age. Our major omission, which we plan to treat in the
future, is the observational selection effects associated with the
detection problem itself.  In effect, the observer's selection function
involves kinematic, geometric and detection-related pieces and we omit
the last, focusing purely on the kinematic and geometric aspects.

In this paper we present a new analysis of pulsar velocities that (1)
tests and exploits the hypothesis that pulsars are born in or near the
plane of the Galaxy; (2) takes into account the uncertainties in
pulsar distances and proper-motion measurements; (3) estimates the
full three dimensional velocity pdf and birth scale height of the
known pulsars; (4) uses the global information to sharpen the estimate
of measured kinematic quantities and to infer the most likely values
of other unmeasured quantities for each pulsar; and (5) constrains the
braking index and torque decay time for pulsar spindown.  Our
methodology uses a likelihood analysis, augmented by a 
comparison of models with different complexity using
Bayesean odds ratios.  This paper is the second paper in a series,
where the first addressed the population of millisecond pulsars,
finding them to be a low-velocity population with rms velocity 
$\sim 84$ \kms (Cordes \& Chernoff 1997; hereafter Paper I).

The first half of the paper deals directly with the statistical
procedures we have developed and the application to the data.  In
\S\ref{sec:pulsar} we describe the selection criteria for the pulsar
sample.  \S\ref{sec:radial} outlines and applies our procedure for
constraining radial velocities and birth altitudes for individual
objects.  \S\ref{sec:v3d} derives a best fit three dimensional
velocity distribution for the population and discuss
the kinematic constraints on the braking index.
\S\ref{sec:veluno}-\ref{sec:pulsars} use the statistical results as an
aid in estimating separate velocity components, birth z and ages for
individual pulsars. Using the estimated velocities we discuss in
\S\ref{sec:vb} the velocity-magnetic moment correlation.
\S\ref{sec:psrsnr} relates recent work on pulsar-supernova remnant
associations to our analysis of the proper motion sample.

The second half of the paper relates our work to a number of topics of
current astrophysical interest.  In \S\ref{sec:astrophy} we review the
lines of evidence for asymmetric supernova kicks and derive the
fraction of the observed sample that will escape the Galaxy. We
discuss the form of the velocity distribution function and give
arguments that suggest that kicks $\simgreat 50$ \kms {\it always} occur
and that a significant fraction of objects are moving so fast that {\it
only} strong supernova kicks can explain their motion. We speculate
about the relationship of the shape of the velocity distribution
function to binary evolution scenarios. We discuss the implications
for core collapse, drawing attention to the breaking of symmetry
implied by the observed kick size.

Finally, we include a discussion of the search for the highest
velocity objects in \S\ref{sec:searching}.  We summarize our results
in \S\ref{sec:summary}.  Individual objects with idiosyncratic
distance estimates or with extreme birth velocities or z altitudes are
discussed in Appendix~\ref{app:individuals}.

\section{PULSAR SAMPLE}\label{sec:pulsar}

The observables we use to analyze pulsar space velocities are the spin
period and its first time derivative ($P$ and $\dot P$); galactic
coordinates ($\ell, b$); the dispersion measure (DM); and the proper
motions in the directions of right ascension and declination ($\mura$,
$\mudec$) and their associated measurement errors ($\errmura,
\errmudec$).  We incorporate auxiliary information on particular lines
of sight that includes the scattering measure (SM) and the locations
of HII regions that may perturb distance estimates based on DM.  The
distance scale is based on the electron density model of TC,
augmented by neutral hydrogen (HI)
absorption measurements and other independent constraints on the
distances, including 2 parallax measurements.  In aggregate, the
available data provide a formal distance estimate and, for
most objects, lower and upper bounds ($D_L, D_U$) such as those
presented by Frail (1990), Frail \& Weisberg (1990) and Koribalski
\etal (1995) based on HI absorption measurements, pulsar associations
with supernova remnants, and parallax measurements.

Table~\ref{tab:psrdata} lists the data that we have used, which
results from the requirement that the spindown age ($P/2\dot{P}$ for
braking index $n = 3$ and no torque decay) be less than 10 Myr and
that the distance estimate include a lower and upper bound. A
chronological age cutoff is necessary because we treat the orbital
motion of the pulsar analytically by a Taylor series approximation and
this ceases to be accurate once significant changes in the
acceleration have occurred.  We focus on the {\it vertical} pulsar
motion in our analysis ignoring all variations with Galactocentric
radius. We choose the age cutoff of 10 Myr to guarantee that virtually
all bound objects are still in the first quarter cycle of their
oscillation perpendicular to the galactic plane. As we discuss in more
detail below, the 10 Myr age cutoff assures quite accurate vertical
solutions in an infinite, planar Galaxy.  However, in 10 Myr a $1000$
km/s pulsar moves $\sim10$ kpc, so the assumption of uniformity in the
plane is nontrivial.

For each object 
the chronological age is based on $P$, $\dot{P}$ and a spindown model. 
It turns out that the
models of greatest interest have $n>3$ and/or torque decay. For them, the
chronological age is less than $P/2\dot{P}$ so the imposed cutoff
of 10 Myr is
more conservative than need be the case. Even for $2 < n < 3$, the
maximum chronological age of 20 Myr degrades but does not destroy the
accuracy of the vertical solution. The age cut implies that our sample
excludes millisecond pulsars and other pulsars that may have undergone
accretion-driven spinup, a process that necessarily is slow and is
therefore manifested in older objects.

Of 96 objects with proper motion measurements, 47 satisfy our age and
distance criteria.  The 49 excluded objects include 45 that are older
than 10 Myr, of which 18 are older than 100 Myr.  An additional 4 objects
were excluded because the TC distance model yielded only a lower bound
on the distance and there was no auxiliary information to provide
better constraints on the distance.   The sample of 47 objects includes
three with parallax measurements 
(B0630+17 [Geminga], Caraveo et al. 1996;
 B0823+26, Gwinn \etal 1986; and
 B2021+51, Campbell \etal 1996).

\section{RADIAL VELOCITIES AND BIRTH ALTITUDES}\label{sec:radial}

We first derive constraints on the radial velocity and birth altitude,
$\zb$, from measurements of proper motion and distances on individual
objects.  In so doing, we demonstrate that all pulsars with measured
proper motions are consistent with birth near the galactic plane,
$\vert \zb\vert\simless 0.3$ kpc.  This kind of analysis was first
applied by Helfand \& Tademaru (1977) to a sample of 12 objects.  Our
conclusion about pulsar birth in the galactic disk is identical to
that of Helfand \& Tademaru. Similarly, we find that the
{\it apparent} motion of a few pulsars toward the galactic plane is due
exclusively to projection effects.  We extend previous analyses
by taking into account errors on estimated distances
and proper motions, acceleration in the galactic
potential and alternative values for the braking index.

In the following,
we review the age uncertainty and describe a model which incorporates
an arbitrary braking index and torque decay. We outline the geometric
relations that exist between the estimated distance, proper-motion
components, z-distance from the Galactic plane, z-value at birth,
radial velocity and pulsar age. We formulate a likelihood function for
a quantity composed of possible z-values at birth and peculiar
radial velocities today. We compare its value on a case-by-case basis
to {\it a priori} expectations.

\subsection{Age Estimate Uncertainties}\label{sec:decay}

The spindown age for an object with period $P$ and period derivative
$\dot P$ is defined as
\begin{equation}
\taus \equiv \frac{P}{(n-1)\dot P}
\label{eq:tau}
\end{equation}
where $n$ is the braking index.  We have $n=3$ for braking by
radiation from a point magnetic dipole of constant strength.  If $n$
is a known constant, $n>1$, and the initial period $P_0 \ll P$ then
the spindown age $\taus$ and the chronological age $t$ agree.

Measured braking indices are smaller than the dipole result: $n=2.28$
for PSR B0540-69 (Boyd \etal 1995); $n=2.51$ for the Crab pulsar
(Lyne, Pritchard \& Smith 1988); and $n=2.83$ for PSR B1509-58
(Manchester \etal 1985; Kaspi \etal 1994).  Most pulsars are much older than
these objects and their braking indices cannot be measured reliably
because timing irregularities mask the
contribution from the small expected second period derivative, $\ddot
P$.
Work below indicates that values of the braking index
$n\simless 3$ (and no torque decay)
are less consistent with the kinematic data than are
values $\simgreat 3$ (and/or torque decays on a time scale of a few
million years). Since most of the pulsars in our sample are
significantly older than those for which braking indices have been
measured directly, these facts imply 
that braking indices are not constant and/or
torque decay is significant over a pulsar's lifetime.

In principle, many physical effects can alter spindown from the static
magnetic dipole limit. The strength of the magnetic field may change:
Ohmic decay (Ostriker and Gunn 1969) and field growth (e.g. by thermal
effects, Blandford, Applegate and Hernquist 1983) are both possible.
Particle emission processes may be more important than dipole
radiation in angular momentum loss (Michel 1969, Goldreich and Julian
1970) and the losses might scale differently with frequency and with
magnetic field than dipole losses. 
The dipole geometry may change in time: some
systematic changes in the relative orientation of the spin axis and
the dipole axis were observationally inferred from the shape of pulsar
beams (Lyne \& Manchester 1988).  The effective size of the dipole may
change in time (Melatos 1997).  Surface fields (as opposed to the
field at the light cylinder) may be altered by mass accretion (Romani
1990) and crustal plate motion (Ruderman 1991); whether such changes
influence the spindown rate is unclear.

It is difficult to test these ideas directly since measurement
of torque decay may be masked by other physical processes occurring
inside the neutron star. Some information is available. General
observational indications are that pulsars that are or were members of
binaries have weak fields of $10^8-10^9$ G that do not decay over
lifetimes of $10^9-10^{10}$ yrs. In a number of recent, specific
analyses involving accretion in binary systems field decay has been
inferred to occur (e.g. Burderi, King \& Wynn 1996).  Current
statistical studies of pulsar properties do not find strong evidence
for field decay in single pulsars (e.g. van den Heuvel 1993) nor
for changes in the relative orientation of spin and dipole axes
(Bhattacharya \& van den Heuvel 1991). Recently, Wang (1997) has
argued that an isolated pulsed X-ray source accreting from the
interstellar medium must have undergone field decay from $\sim 10^{12}$ G to
$10^{10}$ G in $\sim10^7$ yrs (by power law decay) or $\sim10^8$
yrs (by exponential decay).

In view of the uncertainties, we adopt a phenomenological approach and
write $\dot\Omega = -K(t) \Omega^n$, where $t$ is the chronological
age and $K(t)$ describes any time dependence of the torque law other
than a pure power of the spin rate (``torque decay'').  Most
discussions on torque decay refer explicitly to exponential decay of
the magnetic field for $n=3$.  Then $K(t)\propto B^2$, and the quoted
{\it field} decay time is twice that of the torque decay.  The notion
of magnetic field decay has changed considerably since it was first
introduced in discussions of pulsar evolution.  Ohmic decay in the
neutron star crust was initially thought to be fast enough to account
for the turn-off of radio emission from pulsars (Gunn \& Ostriker
1970) but it is now regarded as too slow to be relevant (e.g.
Goldreich \& Reisenegger 1992).  For our uses, the phenomenological
model applies for any $n > 1$ but the relation between $B$ field and
$K(t)$ will vary.

For exponential torque decay, $K(t)\propto\exp(-t/\tau_K)$ and the
chronological age is given by (Bailes 1989)
\begin{equation} t =\tau_K \ell n
    \left \{ 
           1 + \frac{\taus}{\tau_K}
              \left [ 1 - \left ( \frac{P_0}{P} \right )^{n-1} \right ]
    \right \}.
\label{eq:chrono}
\end{equation}
If $t \ll \tau_K$ and $P \gg P_0$, then $t \sim \taus$. In other words,
the spindown age and the chronological age agree when the torque
decay has not taken place and when the initial period is short compared
to the current period. On the other hand, if $t \simgreat \tau_K$ or
if $P \sim P_0$, then $t \ll \taus$ implying
that the spindown age overestimates
the chronological age.

A braking index may be inferred from the instantaneous rate of
change of frequency. The {\it estimated} braking index is
\begin{equation}
\hat n \equiv \frac{\ddot\Omega\Omega}{\dot\Omega^2} = 
     n + (n-1)\frac{\taus}{\tau_K} .
\label{eq:bindex}
\end{equation}
For $\taus \ll \tau_K$, ${\hat n} = n$ but for
$\taus \simgreat \tau_K$ and $n>1$, ${\hat n} > n$.
This sort of evolution would be consistent with both the small measured
values of $\hat n$ for very young pulsars and the larger kinematically
inferred values of $\hat n$ for older objects (see below,
\S\ref{sec:braking}, where we discuss constraints on spindown laws).

\subsection{Coordinates, Kinematic and Peculiar Velocities, \& 
Proper Motions}\label{sec:coord}

We adopt $U,V,W$ coordinates (Mihalas \& Binney 1981, pp. 382-383)
where $U$ and $V$ point toward $\ell = 180^{\circ}$ and $\ell = 90^{\circ}$,
respectively, for $b=0$ and $W$ points toward $b=90^{\circ}$. 
We transform the measured proper motions and their errors to the UVW system
to obtain $\mu_{U,V,W}$ and $\sigma_{\mu_{U,V,W}}$.   

Let $z$ be the present distance from the plane, $\zb$ be the distance
at birth from the plane\footnote{We assume the Sun is at midplane,
even though there is evidence that it currently is $\sim 15$ pc above
the plane (Hammersky \etal 1995).} and $\ddot z$ be the acceleration
at the current position. The present-day, ``kinematic'' $z$ velocity of
an object of age $t$ is
\begin{equation}
V_{z} \approx \frac{z-\zb}{t} + \frac{\ddot z t}{2} 
\label{eq:vzkin}
\end{equation}
and the birth z velocity is
\begin{equation}
{V_{z}}_0 \approx  V_z - {\ddot z}t.
\label{eq:vzbirth}
\end{equation}
The Taylor series expansion is accurate for pulsars younger than $\sim
1/4$ their z-oscillation period. 
Figure ~\ref{fig:vzmodel} shows $V_z(t)$ for pulsars
with $z_0=0$ moving in a three-component model for
the galactic potential (Paczynski 1990).  The very slowest pulsars
(e.g. ${V_z}_0 \simless 10$ \kms), have asymptotic periods of 60-70 Myr
(near the Sun), while somewhat faster pulsars (${V_z}_0 \sim 50$ \kms)
have periods $\sim 100$ Myr.  By restricting our analysis to objects
with $\taus < 10$ Myr, we can safely assume that our expressions above
accurately account for vertical acceleration.
For example, with ${V_z}_0 = 100 $ km s$^{-1}$ at
20 Myr, the Taylor series calculated velocity differs from the true
current velocity by less than 3\% despite changes of order 33\%.
Faster pulsars show smaller relative errors and slower ones, larger
errors. (For initial z velocities of 20 km s$^{-1}$, the error is 4 km
s$^{-1}$ in the final z velocity.)  Since our main purpose is to
describe pulsar velocities statistically over the broad range of
values encompassed by pulsars, this precision is more than sufficient. As a
practical matter, errors in pulsar distances and the proper motions
themselves dominate uncertainties in the approximations associated with the
vertical motion.

The pulsar peculiar velocity (motion with respect to the mean disk in
the vicinity of the pulsar) is
\begin{equation}
{\bf V}^{(P)} = D \muvec^{(P)} + \vr {\bf\hat{n}}
\label{eq:peculiar-velocity}
\end{equation}
where $D$ is the distance, $\muvec^{(P)}$ is the peculiar proper motion and
$\vr {\bf\hat{n}}$ is the line-of-sight peculiar radial velocity.
The proper motion $\muvec = \muvec^{(P)} + \muvec^{(DGR)}$ where
$\muvec^{(DGR)}$ is the known 
contribution from differential galactic rotation (DGR).
The actual radial velocity is also composed of a peculiar part and
a known contribution from DGR ($\vri = \vr + \vrd$).
The z-component of the peculiar velocity is
\begin{equation}
V_z^{(P)}    = D {\mu_W}^{(P)} + \vr \sin b
\label{eq:peculiar-zvelocity}
\end{equation}
and $V_z = V_z^{(P)}$ since the effects
of rotation lie in the plane. One combination of the individually
unmeasurable quantities $\vr$ and $z_0$ is determined in terms of
the distance, current acceleration, peculiar proper motion and age
(assumed known):
\def\vreff{{V_{r,eff}}}
\def\vreffmean{{\overline{\vreff}}}
\begin{equation}
\vreff
\equiv \vr + {z_0 \over t \sin b} = {D \over t} + {1 \over \sin b}
\left( {\ddot z t \over 2} - D \mu_W^{(P)} \right) .
\label{eq:vreff}
\end{equation}
Note that $\vreff$ relates a particular combination of initial birth
location and current peculiar radial velocity to directly measured
quantities.

\subsection{Likelihood Function for Radial Velocities \& Birth Altitudes}
\label{sec:application}

We will use $\vreff$ to characterize individual objects: large values
imply large peculiar radial motions and/or large initial birth heights.
We seek the pdf of $\vreff$ given the uncertainties in the proper
motion measurements and in the distance to individual objects. We treat
this analysis as a preliminary attack on the determination of the
3D velocity distribution which is the subject of the following section.

We write the probability density function for the error in
the proper motion measurement $f_{\delta\muvec}(\delta\muvec)$
where the error is
\begin{equation}
\delta \muvec = \muvec - \data
\label{eq:muw}
\end{equation}
and $\data$ is the reported value. We assume that the pdf for $\delta \muvec$ 
is a Gaussian with reported standard deviation $\errmu$.
For the distribution of
$\vreff$ we are concerned with the component of proper motion
out of the plane. Denote by $\vreffmean$ the value of
$\vreff$ when $\mu_W^{(P)} = \dataw - \mu_W^{(DGR)}$ in Eq.
\ref{eq:vreff} above. The pdf for an object with distance $D$,
in direction ${\hat n}$ and chronological age $t$ is
\begin{eqnarray}
f_{\vreff}(\vreff\,\vert\,\dataw, D, {\hat n},t ) &=& 
               \int d\delta\mu_W f_{\delta\mu_W}(\delta\mu_W)
               \delta (\vreff - \vreffmean - [D\delta\mu_W/\sin b]) 
   \nonumber \\
  &=& {\sin b \over D} 
     f_{\delta\mu_W}(\mu_W^{(P)} + \mu_W^{(DGR)} - \dataw).
\end{eqnarray}
In the second line, $\mu_W^{(P)}$ is given explicitly by
\begin{equation}
\mu_W^{(P)} = {\sin b \over t} + {\ddot z t \over 2 D} - {\sin b \over D}
\vreff ,
\label{eq:murelation}
\end{equation}
where $\mu_W^{(DGR)}$ and $\ddot z$ are the position dependent
differential galactic rotation contribution to the proper motion
and the position dependent galactic acceleration. (For notational
simplification, we suppress writing the direction and age in the pdf's
that follow in this section.)

There is significant uncertainty in the distance to most pulsars.  The
pdf of $\vreff$ is expressed in terms of the range and distribution of
distances allowed by observations. We have
\begin{eqnarray}
f_{\vreff}(\vreff\,\vert\,\dataw,D^*) &=& 
                     \int dD \,f_{D}(D) 
f_{\vreff}(\vreff\,\vert\,\dataw, D) 
\end{eqnarray}
where $D^*$ stands for the imperfect distance knowledge
and $f_D(D)$ is the corresponding pdf for distance.  The specific
form for the distance distribution may not be critical for all
applications. If we are primarily looking for objects that are grossly
discrepant with what we know about the disk scale height and the
typical velocity of pulsars today, then a simple approximation may be
sufficient. We assume a flat distribution for the
distance in the interval $[D_L, D_U]$ where the lower and upper
distance bounds are based on considerations of the TC distance model
along with HI absorption and other measurements. In later sections,
we will use an expression for $f_D(D)$ that is based on the expected
number of pulsars in a population that would be visible.

The pdf has the form
\begin{equation}
f_{\vreff}(\vreff\,\vert\,\dataw,D^*)
 =  \frac{\sin b}{(D_U-D_L)}
               \int_{D_L}^{D_U} dD \, D^{-1} 
      f_{\delta\mu_W}(\mu_W^{(P)}+ \mu_W^{(DGR)} - \dataw).
\label{eq:like1}
\end{equation}
We use it to derive estimates for the covariant combination of the
peculiar radial velocity and birth altitude of a given pulsar.  The pdf
determines the most likely value for $\vreff$ and associated confidence 
interval
[$\vreff^-, \vreff^+$].  Then values for the peculiar radial velocity
$\vr$ and birth height $\zb$ and individual confidence intervals may
be found by using Eq.~\ref{eq:vreff}.

Initially, our analysis uses the chronological age $t = \tau_S$ for
braking index $n=3$ and no torque decay.
Later in the paper, we consider alternative age estimates.
To calculate the accelerations $\ddot z$ in Eq.~\ref{eq:murelation} we
use a three-component model for the galactic potential (Paczynski
1990).  To model differential galactic rotation, we use a flat
rotation curve with circular velocity $v_{rot} = 220$ km s$^{-1}$.
Figure~\ref{fig:zbvr} shows $\vr$ plotted against
$-0.5 \le \zb \le 0.5$ kpc for all the pulsars in the sample. 
Also shown for three pulsars are dashed,
parallel lines that designate the 68\% likelihood range in $\vr$ due
to measurement errors in proper motion and to distance uncertainties.
The figure demonstrates that, for most objects, solutions for $(\vr,
\zb)$ exist with $\zb\simless 0.3$ kpc and $\vert\vr\vert\simless
10^3$ km s$^{-1}$.  Three objects (the Crab pulsar, B0540+23 \&
B0611+22) require $\vert \zb\vert\sim 0.2 - 0.3$ kpc within the
plotted range of $\vr$. The allowed solutions imply
that (1) the pulsar data are
consistent with birth in the plane from Population I progenitor stars;
and (2) no known pulsar younger than 10 Myr with a
proper motion measurement need be born high above the galactic plane.

We also considered alternative relations between spindown and
chronological ages.  The braking index value $n=2.5$ yields a roughly
similar set of solutions but the situation becomes increasingly
extreme as $n$ decreases. For $n=2$ some objects (e.g. 1933+16)
have large $\vreff$, e.g.  $\vert \vr \vert > 5000$ km s$^{-1}$ at
$\zb = 0$ kpc, or $\vert\zb\vert > 0.3$ kpc at $\vr = 0$ km s$^{-1}$.
Figure~\ref{fig:zbvr_3psrs} shows $(\vr, \zb)$ solutions for four
pulsars for braking indices of 2, 2.5, 3 \& 4.

\section{THE 3D VELOCITY DISTRIBUTION}\label{sec:v3d}

\def\vb{{V_0}}
\def\vbvec{{ {\bf V}_{\rm 0} }}
\def\vbvecp{{ {\bf V}_{\rm 0}^{(P)} }}
\def\xb{{ {\bf X}_0 }}
\def\age{{t}}

By combining the radial velocity constraints of \S\ref{sec:radial}
with proper motion measurements, it is possible to derive
information on the full 3D velocities of  pulsars and their
birth distances from the galactic plane.  Here we derive the joint
pdf for birth $\zb$ and birth velocity.

We first relate the initial distribution function for pulsars
at birth to the observed distribution function today. Then,
we derive a likelihood function for proper motion data to
be used to infer information about the initial peculiar velocity 
and birth height above the plane.

\subsection{Birth \& Present-day Distribution Functions}

The distribution function for particles moving in a fixed background
potential is related directly to the distribution function of
initial conditions. Choose an inertial frame. At time $t$ let the
position be $\bf X$, the velocity $\bf V$ and let the corresponding
initial conditions be $\xb$ and $\vbvec$.  Liouville's theorem states
that the final and initial
distribution functions satisfy $f_{\bf X, V}({\bf X, V}) = f_{\xb,
\vbvec}(\xb, \vbvec)$. 

We will apply this result to relate the initial and final pulsar
distributions, bearing in mind that $\vbvec$ 
is the sum of the circular rotation velocity and the peculiar
motion, including the birth kick.  The
distribution of initial conditions $f_{\xb, \vbvec}(\xb, \vbvec)$ is
expressed in terms of the initial peculiar velocity
distribution $f_{\xb, \vbvecp}(\xb, \vbvecp)$.
If its age is
sufficiently small, then the equations of motion show
that the changes in peculiar velocity are
small. To be more precise, we ignore terms of {\sc O}$(\Omega t, V_0^{(P)}
t/ R)$ compared to $V_0^{(P)}/v_{rot}$ where $\Omega$ is the circular
frequency, $R$ is the Galactocentric radius and $v_{rot}$ is the
rotation velocity.  We also ignore variations with
$R$, so we factor $f_{\xb,\vbvec} = f_{gp} \times f_{z,\vbvec}(z, \vbvec)$ 
into a constant two-dimensional, galactic plane part ($f_{gp}$) 
and a four-dimensional pdf in $z$ and $\vbvec$.
This is justified for particles (neutron stars)
whose total radial displacement is small compared to the length scale for
variations of disk properties.  This approximation may be problematic
given the maximum distances traveled in 10 Myr.  The physical
content of these assumptions is that the birth rate, birth 
scale height and initial velocity kick distribution do not vary
appreciably over the range of the radial motion, at least
for parts of the Galaxy that contribute to the observed pulsar
sample. We find from
Eqs.~\ref{eq:peculiar-velocity} and \ref{eq:peculiar-zvelocity} that
\begin{eqnarray}
z_0  =  z - V_z t + {1 \over 2} \ddot z \age^2 
   =  z - ( D\mu_W^{(P)} + \vr \sin b) \age + {1 \over 2} \ddot z \age^2
\label{eq:xx}
\end{eqnarray}
\begin{eqnarray}
\vbvecp  =  {\bf V}^{(P)} - {\bf \hat z} \ddot z \age 
         =  ( D \muvec^{(P)} + \vr {\bf\hat{n}} ) - \zhat \ddot z \age .
\label{eq:vv}
\end{eqnarray}
The pdf of present-day pulsars is related to the pdf
at birth by
\begin{equation}
f_{{\bf X,V}}( {\bf X}, {\bf V}) =  
f_{\xb, \vbvecp}(\xb, \vbvecp) =
f_{gp} f_{z_0, \vbvecp} ( z_0, \vbvecp ) ,
\label{eq:link1}
\end{equation}
For convenience, we abbreviate 
$f_{z_0, \vbvecp} ( z_0, \vbvecp ) = f( z_0, \vbvecp )$.

\subsection{PDF of Proper Motion}

\def\vpsr{{\bf V}}
\def\rpsr{{\bf X}}

\def\likeli{{\cal L}_{\data}(\muvec)}
\def\w{{\bf (V - V_{\rm \odot} ) }}
\def\wperp{{\w_{\perp}}}

Now we derive the likelihood function for the proper motion.
We begin by forming the likelihood of a perfectly measured proper 
motion $\muvec$ for a pulsar with perfectly known position $\rpsr$,
displaced from the observer's position by $D \nhat$. The pdf is
\begin{equation}
f_{\muvec} ( \muvec | \rpsr ) = 
   {1 \over {\cal N}_{\muvec\vert \rpsr }}
\int d^3 \vpsr f_{\rpsr, \vpsr} ( \rpsr, \vpsr )
\delta^2 (\muvec - {\wperp \over D}) .
\end{equation}
Here $f_{\rpsr,\vpsr} ( \rpsr, \vpsr)$ is the present-day
pulsar pdf; 
${\cal N_{\muvec \vert \rpsr}} 
 \equiv \int d^3 \vpsr f_{\rpsr, \vpsr} ( \rpsr, \vpsr )
   = f_{\rpsr}(\rpsr)  $
normalizes the proper-motion pdf; 
and $\vsun$ is the velocity of the Sun due to galactic rotation.

The distribution of the {\it measured} proper motion $\data$, 
including measurement errors, is
\begin{equation}
f_{\data}( \data | \rpsr ) = \int d\muvec f( \data | \muvec ) 
f_{\muvec}(\muvec | \rpsr ) ,
\end{equation}
where $ f( \data | \muvec )$ 
is the likelihood of obtaining the measured value
given the true one and is simply the usual Gaussian function
for each component 
with standard deviations that reflect measurement uncertainties. 
From our previous discussion, we assume
$f ( \data | \muvec )$ depends only on the 
difference, $\delta \muvec \equiv \muvec - \data$, written as
$f_{\delta \muvec}(\delta \muvec)$. Let $D^*$ stand
for lower and upper cutoffs for $D$ based on dispersion
measure, known HII regions and so forth. Let
$f_D( D | D^*, \nhat )$ be the resultant probability distribution.
Marginalizing over the true distance,
\begin{equation}
f_{\data}( \data | D^*, \nhat ) = \int dD f_{\data} ( \data | \rpsr )
f_D( D | D^*, \nhat ) .
\end{equation}
We can write the pdf for distances in
the pulsar population described by $f_{\rpsr,\vpsr}$
{\it assuming all such objects are visible}. We find
\begin{equation}
f_D( D | D^*, \nhat ) = { D^2 f_{\rpsr}( \rpsr ) \over
\int dD D^2  \,\, f_{\rpsr}({\rpsr}) } ,
\label{eq:fD}
\end{equation}
where 
the integral is limited to the range implied by $D^*$.
This is the point
where we have ignored the detection-related aspects of the pulsar problem.
Integrating over the perpendicular velocity we find
\begin{eqnarray}
f_{\data}( \data | D^*, \nhat )  = 
\left [
{\cal N}_{\muvec\vert D^*,\nhat}^{-1}
\int dD D^4  \int d\muvec \int dV_r f_{\delta \muvec}(\data-\muvec)  
f_{\rpsr,\vpsr}( \rpsr, \vpsr )
\right ]_{ \vpsr  =  \vsun + D \muvec + \vri \nhat} 
\end{eqnarray}
where $V_r$ is the radial velocity in the inertial frame
and ${\cal N}_{\muvec\vert D^*,\nhat}^{-1}$
normalizes the pdf.
To carry out the
integrations ($d\muvec dV_r$) we change variables as follows:
$\muvec^{(P)} = \muvec - \muvec^{(DGR)}( \nhat, D )$ 
and $\vr = V_r - V_r^{(DGR)}( \nhat, D)$ and use Eqs. 
\ref{eq:xx}-\ref{eq:link1} to give
\def\fmuarg{{ \data - \muvec^{(P)} - \muvec^{(DGR)} }}
\begin{equation}
f_{\data}( \data | D^*, \nhat, t) =
{\cal N}_{\muvec\vert D^*,\nhat}^{-1}
\int dD D^4  \int d\muvec^{(P)} \int d\vr 
f_{\delta \muvec}(\data - \muvec^{(P)} - \muvec^{(DGR)})
  f( z_0, {\bf V}_0^{(P)} ) 
.
\label{eq:pdf_mu}
\end{equation}
Note that $\muvec^{(P)}$ and $\vr$ are dummy integration variables
and that the transformation to the initial peculiar velocity distribution
requires knowledge of the chronological age.
The data and the DGR contribution to the proper motion occur only in the
pdf that describes the error distribution in the form of the
difference: $\data - \muvec^{(DGR)}$. 
For cases of interest ($\taus \simless 10$ Myr),
the acceleration correction to the velocity argument in Eq.\ref{eq:vv}
is only a few tens of \kms and is, therefore, unimportant for establishing
the velocity distribution of high-velocity objects. 
However, the correction
to the spatial argument in Eq.~\ref{eq:xx}
is significant.  In the following, we ignore the
velocity correction but apply the spatial correction.

\subsection{Parameterization \& Likelihood Function}\label{sec:like3d}

The likelihood function for the parameters is the product over
$N_{psr}$ pulsars,
\begin{equation}
{\cal L} = \prod_{k=1}^{N_{psr}} {\cal L}_k,
\label{eq:like_g}
\end{equation} 
where the likelihood factor, 
${\cl_k}\equiv f_{\data}(\data_k \vert {\bf \hat n}_k, D_k^*, t_k)$, 
for each pulsar is simply Eq.~\ref{eq:pdf_mu}
evaluated using the measured proper motions (and errors) along with
the distance constraints $D_L, D_U$, the direction $\hat {\bf n}$
and the age $t$.

To apply Eq.~\ref{eq:pdf_mu},
we adopt  a parametric approach, where we assume particular forms for the 
pdfs in
$\zb$ and ${\bf V}_0^{(P)}$.   Specifically, we assume that $\zb$ 
and ${\bf V}_0^{(P)}$ have a multicomponent Gaussian pdf of the form
\begin{equation}
f (\zb, {\bf V}_0^{(P)}) = \sum_{j=1}^{n_g} w_j  
       g_{1d}(\zb, {h_z}_j)  g_{3d}( {V_0^{(P)}}, {\sigma_V}_j).
\label{eq:pdfs}
\end{equation}
In Eq.~\ref{eq:pdfs}, $g_{1d}(p,\sigma)$ is a standard
1D Gaussian pdf with zero mean and standard deviation $\sigma$,
$g_{1d}(p,\sigma) = (2\pi\sigma^2)^{-1/2}\exp(-p^2/2\sigma^2)$,
while $g_{3d}$ is a 3D Gaussian
$g_{3d}(q,\sigma) =  (2\pi\sigma^2)^{-3/2}\exp(-q^2/2\sigma^2)$.
The weights $w_j$ sum to unity.    
The pdf is defined so that $f dz_0 d^3V_0^{(P)}$ is 
the infinitesimal probability.

We have chosen a multicomponent Gaussian model because its analytical
properties allow it to fit a wide range of shapes for the actual
distributions in $z_0$ and ${\bf V}^{(P)}$.  There is not necessarily an  implied
{\it physical} basis for this choice of form: the different Gaussian
velocity components need not correspond to different population 
components.  

\subsection{Results \& Comparison of Models Using Odds Ratios}\label{sec:odds}

The parameters to be determined for a pulsar sample are
(1) $n_g$ standard deviations for velocities, ${\sigma_V}_j$;
(2) $n_h \le n_g$ scale heights for the birth altitude, ${h_z}_j$; and
(3) $n_g-1$ weights, $w_j$
for a total of $n_g+n_h-1$ parameters. 

We considered a set of models with increasingly complex velocity and
birth height distributions. We label the models by $n_g.n_h$ and briefly
describe them below:
\begin{itemize}
\item[1.] [Model 1.1]:
     a single component model 
    ($n_g = n_h = 1$) 
\item[2.] [Model 2.1]:
     a two component velocity model with a single scale height 
    ($n_g=2, n_h = 1$) 
\item[3.] [Model 2.2]: 
     a model with two scale height and two velocity components 
    ($n_g=2, n_h = 2$) 
and 
\item[4.] [Model 3.1]:
     a three component velocity model with a single scale height 
    ($n_g=3, n_h = 1$) 
\end{itemize}

We assume flat priors for the parameters in selected ranges that
are listed in Table~\ref{tab:ranges}. 
We generated the posterior
probability distribution of parameters which is just the
evaluation of the likelihood in the selected ranges.
The modes of the posterior, i.e. the
maximum likelihood results, and $\log {\cal L}$ appear in Table
\ref{tab:compare}. These are the ``best'' values of parameters for each
model. The analysis was repeated for several age cuts for the
sample: $\taus_{max} =$ 1, 5 \& 10 Myr (for $n=3$ and no torque
decay).  The mode of the distribution is typically
well-defined. For example, in Figure~\ref{fig:contour1.1} we display
contours of $\log {\cal L}$ (log likelihood) for Model 1.1 with
$\taus_{max} = 10$ Myr. The mode is identified by the cross. In Figure
\ref{fig:contour2.1} we display contours for Model 2.1 in two-dimensional
surfaces that pass through the mode. 

The results in Table~\ref{tab:compare}
clearly suggest the presence of a significant
high velocity component when the models are sufficiently complex.  At
the same time, examination of the likelihood contours in the simplest
model does not provide any obvious indications
that its single $300$ km s$^{-1}$ component is
inadequate. To make such a determination, we must compare the models
in a systematic manner.

To compare models we use the odds ratio (cf. Gregory \& Loredo 1992) to
quantify goodness of fit while penalizing models with more parameters. 
We assume that all models are equally probable, {\it a priori}.
The odds ratio reduces to the ``Bayes factor,'' which is the ratio of
global likelihoods of two models (Gregory \& Loredo's Eq. 2.12). 
The global likelihood for a model M given data $\cd$ is
\begin{equation}
f(\cd\vert M) = \int d\thetavec f_{\thetavec}(\thetavec\vert M)
   \cl(\thetavec),
\label{eq:gl}
\end{equation}
where $f_{\thetavec}(\thetavec\vert M)$ is the prior pdf for model parameters
$\thetavec$ and $\cl(\thetavec)$ is the likelihood function as we have
used it in this paper. We have already assumed the prior pdf is flat and
we have seen that $\cl(\thetavec)$ is strongly peaked.
Thus, letting $\hat\thetavec$ be the parameters that maximize
$\cl(\thetavec)$,  
\begin{equation}
f(\cd\vert M) \sim 
f_{\thetavec}(\hat\thetavec\vert M)
\int d\thetavec \cl(\thetavec) \equiv \frac{\Lambda_M}{V_M},
\label{eq:gl2}
\end{equation}
where the last equation defines the integrated likelihood, 
$\Lambda_M \equiv \int d\thetavec\cl(\thetavec)$,
and the volume in parameter space that is searched, 
$V_M = \left[ f_{\thetavec}(\hat\thetavec\vert M)\right ]^{-1}$.

We take the single Gaussian model with two parameters as our reference model:
$M_1 \equiv 1.1$.
The odds ratio for the M$^{th}$ model relative to $M_{1.1}$ becomes
\begin{equation}
O_{M,M_1} \equiv \frac{f(\cd\vert M)} {f(\cd\vert M_1)}
\equiv 
\frac{\Lambda_M}{\Lambda_{M_1} }
\frac{V_{M_1}}{V_{M} },
\label{eq:odds1}
\end{equation}
which we evaluate through numerical integration of the likelihood function 
for each model over a uniform grid in parameter space with bounds given
in Table~\ref{tab:ranges}. The associated
volumes for the four models are:
\begin{eqnarray}
V_{1.1}  =  \Delta {h_z}_1 \Delta{\sigma_V}_1 \nonumber \\
V_{2.1}  =  V_{1.1}\,\,\Delta w_1 \Delta{\sigma_V}_2  
            A_{2.1}\nonumber \\
V_{2.2}  =  V_{1.1}\,\,\Delta w_1 \Delta{\sigma_V}_2 \Delta {h_z}_2  
            A_{2.2} \nonumber \\
V_{3.1}  =  V_{1.1}\,\,\Delta w_1 \Delta {w_2} \Delta{\sigma_V}_2\Delta{\sigma_V}_3
            A_{3.1}
\end{eqnarray}
where $A_{2.1}, A_{2.2}$ and  $A_{3.1} \le 1$ are
factors that account for the overlap in our search of parameter space
for the maximum likelihood and the resulting degeneracy of the 
models.

The maximum likelihood values and the volumes yield the odds ratios
given in Table~\ref{tab:compare}. The odds ratios show that the data
strongly favor the multiple-component models (models 2.1, 2.2 and 3.1)
over the single-component model (1.1). 
However, the additional
complexity of models 2.2 and 3.1 over 2.1 is not demanded by the data.
This conclusion holds for the two largest age cutoffs considered (5
and 10 Myr).  For a 1 Myr cutoff, the small number of data points
appears to reduce the discriminating power between models and, consequently,
we have ignored these results. 
The model fitting suggests that a single birth-z
scale height of 0.13 kpc applies to both velocity components, which
have rms velocities (3D) $\sim 175$ km s$^{-1}$ and $\sim 700$ km
s$^{-1}$.

To refine the estimation, we derive
the marginal distribution of each parameter.  For a given parameter
$\theta_j\in\thetavec$, the marginal pdf is
the normalized integral over all other parameters
\begin{equation}
f_{\theta_j}(\theta_j) =
   \frac{
       \int_{exc. \theta_j} d\thetavec {\cal L}(\thetavec) 
        }
        {
       \int d\thetavec {\cal L}(\thetavec)
        },
\label{eq:thetamarg}
\end{equation}
where the integral subscript `exc. $\theta_j$' means that all parameters
except the j$^{th}$ one are integrated over. Fig~\ref{fig:marg}
illustrates the marginal distributions for Model 2.1.
Table~\ref{tab:bestfit} gives the best-fit values of parameters for
the preferred 
Model 2.1 along with 68\% confidence intervals.  The confidence interval
was calculated by finding the region in each marginalized
distribution  containing 68\% of the area.

The mean 1D, 2D and 3D speeds are given by 
$\langle V_{1D} \rangle = \sqrt{2/\pi} \sum_j w_j {\sigma_V}_j$,
$\langle V_{2D} \rangle = \sqrt{\pi/2} \sum_j w_j {\sigma_V}_j$,
$\langle V_{3D} \rangle = \sqrt{8/\pi} \sum_j w_j {\sigma_V}_j$.
For the best fit model (2.1),
$\sum_j w_j {\sigma_V}_j \approx 264$ \kms and the three mean
speeds are 211, 331 and 421 \kms.  The speeds for the three-component
Gaussian model (3.1) are nearly identical, while they are
about 14\% larger for the single-Gaussian model (1.1).

\subsection{Spindown Ages \& the Braking Index}\label{sec:braking}

The likelihood analysis we have presented is based on a braking index
$n=3$, an infinite torque decay time, $\tau_K$ (cf.
Eq.~\ref{eq:chrono}), and a negligible birth period, $P_0$.  The
analysis is sensitive to the spindown model because we infer the
chronological age $t$ from the spindown age $\tau_S$.  For example,
Eq.~\ref{eq:pdf_mu} shows that the likelihood function depends on the
initial distribution at $\zb$ and Eq.~\ref{eq:xx}, in turn, shows that
the birth position $\zb$ is a direct function of $t$ as well as other
observed ($z$) and marginalized ($\vr$, $\mu_W^{(P)}$) quantities.
Here, we investigate how the results for the 3D velocity distribution
depend on the braking index and decay time.  As mentioned earlier,
measured braking indices for very young pulsars (age $\sim 10^3$ to
$10^4$ yr) are less than 3, yielding spindown ages that are larger
than we have assumed in our use of the conventional spindown age with
$n=3$. As $n$ and $\tau_K$ are varied, ages for some objects increase
while others decrease.

Figure~\ref{fig:ntau} shows the log likelihood for the double Gaussian
model (2.1) plotted against torque decay time, $\tau_K$, for various
values of braking index. {\it All} model parameters were varied to
find the maximum likelihood. The likelihood peaks at $(n,\tau_K)
\approx (2.5, 3 \,\, {\rm Myr})$ and the range of variation in the
final likelihood value is small compared to that due to 
other model parameters.  For example, the likelihood at
$(n,\tau_K) = (3, 5 \,\, {\rm Myr})$ is only 22\% smaller than the
peak.  If torque decay does not occur (i.e. $\tau_k\to\infty$), the
peak likelihood is found at $n\approx 4.5$, but with a likelihood 
that is a factor 20.3 smaller than for $(n,\tau_K) \sim (2.5,
3\,\, {\rm Myr})$.  If torque decay is due to magnetic field decay
with $n=3$, the field decay time is twice the torque decay time, or
$\sim 10 $ Myr.

The best values
for the birth scale height, velocity parameters, and relative weights
do not change appreciably with braking index and decay time. 
(It should be pointed out that we have used a fairly coarse grid 
in this analysis: 
$\Delta w_1 = 0.1$, 
$\Delta {h_z}_1 = 0.05$ kpc,
$\Delta {\sigma_V}_1 = 50$ \kms and
$\Delta {\sigma_V}_2 = 200$ \kms.)
On the other hand,
the mean sample age does
vary in a easily characterized way. It is 2.95 Myr for
$(n,\tau_K)=(3,\infty)$.  For the best fit with $(n,\tau_K) =
(2.5,3\,\,{\rm Myr})$ it is 1.64 Myr while the best fit with no torque
decay for $n=4.5$ yields 1.68 Myr.  In the case with no torque decay,
the best fit reduces all age estimates by the same factor, $3/4.5 =
0.67$.  By contrast, for the absolute best fit for $(n,\tau_K) =
(2.5,3\,\,{\rm Myr})$, some age estimates are increased while others
are decreased.

Even though all model parameters were varied, the best parameter
values for $w_1, {h_z}_1, {\sigma_V}_1$ and ${\sigma_V}_2$ do not
change significantly as the braking index and torque decay time 
are varied. Certain model
parameters (e.g. $\sigma_V$) are directly tied to a measured quantity
($\muvec^{(P)}$) and, hence, not susceptible to changes in t. Other
model parameters (e.g. $h_z$) are strongly influenced by young objects
at small $\sin b$ that have not had the opportunity to move far for
any plausible ages.  In addition, there are substantial errors in
distance and proper motion measurements and these overlay the
intrinsic timescale change.  Typical distance errors are factors
of $\sim 2$ and typical proper motion errors are $\sim 50\%$. Thus
it is not unreasonable that we
distinguish spindown models that have typical age estimates that
differ by $\sim 2$.

The pulsar sample we have analyzed is more consistent
with ages that are smaller, on average, than the conventional spindown
ages.  If no torque decay occurs, then ``large'' braking indices,
$3\simless n \simless 6$ are preferred over small braking indices,
such as $n\sim 2.5$ measured for the Crab pulsar and $n\sim 2.4$ for
B0540-69 in the LMC.  If torque decay occurs, as is suggested by our
best fit model, then a braking index of $2.5-3$ is preferred.  An
average braking index as small as 2 yields a likelihood that is $\sim
10^{2.9}$ times worse than the best fit model.

Our results are therefore at odds with those of Lyne \etal (1996), who
estimate a braking index $n\sim 1.4\pm 0.2$ for the Vela pulsar and
who speculate that braking indices decrease as they age, at least in
going from Crab-like pulsars $\sim 1000$ yr old to Vela-type pulsars
about 20 times older.  Lyne \etal conclude that pulsars are generally
older than the conventional spindown age, in conflict with our
conclusion.

Assuming the braking index for the youngest objects has been
accurately determined, our results could be consistent with
one of several proposed
scenarios in which the braking index increased with age. Finite size
corrections to the dipole spindown model (Melatos 1997) predict such
an evolution.  Heintzmann \&
Schrueffer (1982) noted that braking indices of older pulsars
should be 3 or more based on plasma effects in the magnetosphere.
Muslimov \& Page (1996) suggested that, if the crustal
magnetic field of neutron stars grows at  early times (0.1 -
10 kyr), then braking indices should increase from $n<1$ and asymptote to
$n\to 3$ on a time scale of $\sim 10$ kyr.

In this analysis we have assumed that the birth spin period is much
smaller than the present day period.  There is evidence of
discrepancies between spindown ages of some millisecond pulsars and
ages constrained by temperatures of their white-dwarf companions (e.g.
Camilo \etal 1994).
Vivekenand \& Narayan (1981) and Phinney (1996,
private communication) have argued that pulsars may be born with
nonnegligible spin periods (``injection''), in which case spindown
ages overestimate chronological ages. The fact that the mean sample
age of our most likely models is less than that of the magnetic dipole
model without torque decay would be consistent with this suggestion.
However, it does not follow that our analysis favors injection. In
varying the spindown laws, the ages of all objects change, whereas if
pulsars have a range of birth spin periods only objects with current
periods comparable to birth periods have significantly altered ages.
The set of spindown models we have tested does not cover the
hypothesis of injection.

In \S\ref{sec:ages} we explore the issue of pulsar ages in
generality by deriving the distribution of chronological age
for individual pulsars. This treatment starts from the kinematic data
for each pulsar and our inferred scale height and velocity dispersion
to deduce the age. The treatment is independent of the spindown model,
{\it in so far as the scale height and velocity dispersion parameters
are themselves insensitive to the precise form of the spindown model},
as appears to be the case.  The general conclusion of the case-by-case
analysis is consistent with the above conclusions: pulsars are on
average roughly 50\% younger than their conventional spindown ages.

\section{3D VELOCITIES OF INDIVIDUAL OBJECTS}\label{sec:veluno}

In this section we use the global properties of the observed pulsar
population (from \S3) to aid estimation of the properties of
individual pulsars.  First, we write out a formal expression for the
pdf for ${\bf V}^{(P)}$ and $z_0$ for an individual object given its
distance limits, proper motion measurements, and the chronological age. In
constructing the pdf, we incorporate as prior information the
previously derived parameters for the distribution of birth altitude
and velocity.  Second, we Monte Carlo the resulting pdf to determine mean
and mean square values for each pulsar in the sample.
This highlights a number of interesting objects.

The analysis here is similar to the analysis of \S\ref{sec:radial} but differs
in that we estimate transverse speeds for each object and incorporate
what we have learned from the 3D analysis of the whole population.
The 3D velocities of individual objects can be used to test for
correlations with other pulsar parameters.

Let $Q$ be a quantity whose pdf we seek; $Q$ may be a function of any 
of the observables, with functional form ${\tilde Q}$.
 Signify the data for a given
pulsar as ${\cal D} \equiv \{ D^*, \data, \nhat\}$.
For each object we estimate the chronological age 
$t$ from $P, \dot{P}$ and an assumed spindown model. 
Let ${\cal I}$ represent the
background information, including the posterior probability
distribution for the model parameters governing the birth scale
height, the velocity kick distribution and the relative weights of
components.  Since the posterior was found to be strongly peaked at
the mode and the mode insensitive to the spindown model, we simply
assume that ${\cal I}$ sets the model parameters at the mode (i.e.
the most likely values). We derive
$f_Q( Q | {\cal D}, t, {\cal I})$ following the same general procedure used to
derive Eq.~\ref{eq:pdf_mu}.  First, form the pdf for $Q$ in terms of
the distribution function for today's pulsar positions and velocities,
assuming no observational errors in $\data$ and distance.  Second,
marginalize over the distance range and the proper motion errors. As
before, we take Eq.~\ref{eq:fD} for the pdf for distance $D$ between the upper
and lower limits.  Third, make the change of variables $d^3{\bf V} \to
D^2 d\muvec^{(P)} d\vr$ and replace the present day pulsar
distribution function with the corresponding initial
distribution function. The result is 
\begin{equation} f_Q( Q | {\cal D}, t, {\cal I} ) = 
{\cal N}_{Q}^{-1}
 \int dD D^4 \int d\muvec^{(P)} \int dV_r 
f_{\delta \muvec} (\fmuarg) 
f (z_0, {\bf V}_0^{(P)})
\delta ( Q - {\tilde Q} )
\end{equation}
where ${\cal N}_{Q}^{-1}$ is the normalization constant;
$z_0$ and ${\bf V}_0^{(P)}$ are the initial conditions (given by
Eqs.~\ref{eq:xx}
and \ref{eq:vv});
 and the range of $D$ integration is determined by
$D^*$.  
The previously derived 
parameter values for birth scale height and peculiar velocity 
and the age occur in $f(z_0, {\bf V}_0^{(P)})$.

As a specific example of immediate interest, 
we take $Q = \{ z_0, {\bf V}^{(P)} \}$ to give
the joint pdf for the present day peculiar velocity and the birth
altitude of an individual pulsar
\begin{eqnarray}
f_{z_0, {\bf V}^{(P)} }( z_0, {\bf V}^{(P)} ) & \propto &
\left [
 D^2 f(z_0, {\bf V}_0^{(P)}) 
f_{\delta \muvec} (\fmuarg) 
| {dD \over dz_0 } | 
\right ]_{\displaystyle D = {\hat D}} \nonumber \\
{\hat D} & = &
 {z_0 + \zhat \cdot {\bf V}^{(P)} - {1 \over 2} \ddot z t^2
   \over
  \sin b
 } .
\label{eq:pdf_zv}
\end{eqnarray}
The solution for $\hat{D}$ is implicit because the vertical
acceleration depends on $D$.

The expression clearly shows the role of the initial birth
and velocity distribution $f(z_0, {\bf V}_0^{(P)})$
 on limiting the allowed range of
properties for a pulsar observed today.   For poor or no
observational constraints on proper motion, the birth distribution
functions as a prior.  However,  
if a proper motion measurement
is precise, then $f_{\delta \muvec}$ provides a strong, perhaps
dominant constraint on the velocity components and birth $z$ for the
object. 

To apply Eq.~\ref{eq:pdf_zv} for each pulsar,
we generate 200 Monte Carlo samples
of $z_0$ and ${\bf V}^{(P)}$ using the method of rejection
(cf. Rubinstein 1981; Press \etal 1992).
For the pulsar age, we have assumed the canonical
age given by Eq.~\ref{eq:tau} with $n=3$ 
(i.e.  magnetic dipole radiation and no torque decay).
Table~\ref{tab:components}
gives the mean and rms values for 
$V\equiv \sqrt{\vr^2 + \vert \vpmath \vert^2}$,
$\vr$, 
$\vpmath$,
and $z_0$ for the full sample.
The two components for 
the perpendicular velocity ($V_{\perp 1}$, $V_{\perp 2}$) 
given in the table correspond to 
components in the right-ascension and declination directions.
The rms values indicate the
spread in possible solutions for each variable, subject to 
measurement errors in the proper motion, distance estimation
errors and the fact that only two velocity components are determined
by proper motion. 
Most objects show small average values for
$\vr$, but with a large spread due to uncertainties in
distance and proper motion.  On the other hand, 
three objects require significant radial velocities, 
$\vr > 3{\sigma_\vr}$: B0943+10, B1953+50 and B2154+40.   The latter
two objects require radial velocity magnitudes
$\sim 1500$ and $\sim 1000$ \kms, respectively. 
There  are 8 objects with 3D mean velocity estimates exceeding 500 \kms:
B1508+55, B1706-16, B1842+14, B1933+16, B1946+35, 
B1953+50, B2154+40 and B2224+65.
Of these, five  have velocities in excess of 1000 \kms.\ 
Finally, 9 objects have birth $z$ values that are determined
to better than $3{\sigma_{z_0}}$: B0136+57, B0531+21 (Crab), 
B0540+23, B0611+22, B0630+17 (Geminga), B0656+14, B0740$-$28,
B0833$-$45 (Vela), and B1929+10.  
Of these, three pulsars have $\vert\zb\vert \simgreat 0.2$ kpc.
It is notable that these high $\zb$ pulsars
are in the galactic anticenter.  

\section{PULSAR AGES}\label{sec:ages}

Now we derive the pdf for the chronological age, $t$, of a pulsar given its
proper motion and distance measurements.  Previously, 
we estimated $t$ from the spindown time $\tau_S$ or its variants
using alternative braking indices and torque decay times.
Here, we rely solely on the kinematic
information associated with each object and with the global
 parameters that characterize the birth scale height and 
initial peculiar velocity.
These parameters were determined for a range of spindown models in
\S\ref{sec:braking} and the results were found to be only weakly
dependent on breaking index and torque decay time. This simplification
means that the pdf for $t$ is insensitive to the spindown
assumptions made in the inferring the scale height and peculiar
velocity parameters.

As we have previously emphasized, 
uncertainties in position and velocity
are inextricably linked to the age uncertainty. We use Bayes'
Theorem to relate the age pdf to the likelihood function for the
data ${\cal D} = \{ \data, \nhat, D^* \}$:  
\begin{equation}
f( t | {\cal D, I} ) f( {\cal D} | {\cal I} ) =
f( {\cal D} | t, {\cal I} ) f( t | {\cal I} ) .
\end{equation}
We assume that the prior probability for $t$, $f(t | {\cal I})$,
is constant, consistent with a constant neutron-star birth rate
over the recent past.  We evaluate the likelihood of the
data in the same manner as Eq.~\ref{eq:pdf_mu} to find
\begin{equation}
f_t(t) = f(t | {\cal D, I} ) = 
{\cal N}_t^{-1}
 \int dD  D^4 \int d\muvec^{(P)} \int  d\vr 
f (z_0, {\bf V}_0^{(P)})
f_{\delta\muvec}(\fmuarg) 
\label{eq:pdft}
\end{equation}
where ${\cal N}_t$ is the normalization constant.

Application to individual pulsars is straight forward.  
Figure~\ref{fig:agepdfcdf} shows the pdfs $f_t(t)$ and corresponding cdfs
for three pulsars.  Also shown in the figure, as vertical lines, are
the conventional spindown age estimates, $\taus$.
The shapes of the individual pdf's are directly related
to the Gaussian shapes of the assumed underlying pdf's for $z_0$,
${\bf V}_0^{(P)}$ and $\dmuvec$.  By inspection of Eq.~\ref{eq:pdft},
if measurement errors are infinitesimal and the values for birth $z$
and $D$ are fixed, then $f_t(t)$ has weight where $f (z_0, {\bf
V}_0^{(P)})$ maximizes, in particular, near the origin.  If acceleration is
not significant then the minimum $| {\bf V}_0^{(P)} |$ implies $\vr =
0$ and the minimum $z_0$ implies $t_0 \sim \sin b / \muw$. Thus, the
maximum (or mode) of $f_t$ is at an age $t_0$. The figures show that
the resultant pdf is not symmetric, falling off more slowly toward
large $t$ than small $t$ because of the effect of peculiar radial
velocity and the assumed Gaussian form: 
large $\vr$ (giving small $t$) has exponentially small
probability of occurrence whereas small $\vr$ (giving large $t$) has
only algebraically small probability. 

Figure~\ref{fig:ages} shows the 68\% probability ranges for the ages
of the 47 pulsars in our sample along with the median, mode,
and spindown age.  As demonstrated in
\S\ref{sec:braking}, our likelihood analysis shows a better fit to the
pulsar data for combinations of braking index and torque decay time
that make ages of most pulsars {\it smaller} than the conventional
spindown age, $\taus = P / 2\dot P$.  The conventional spin down time
assumes a braking index of 3, an infinite torque-decay time, and birth
period much smaller than the present-day period. The same trend is
apparent in Figure~\ref{fig:ages} which shows that the majority of objects
have spindown ages $\taus > t_0$.

To further demonstrate that a bias exists between chronological
and spindown ages, we
inspect the ratio of chronological age to spindown age,
$r = t / \taus$. The pdf for this ratio is
\begin{equation}
 f_r(r) = \taus f_t(r\taus).
\end{equation}
We average  $f_r(r)$ over 45 objects in our sample,
excluding the Crab and Vela pulsars because their
kinematic ages are poorly constrained. 
Figure~\ref{fig:sumpdfages} shows the average
$f_r$, which peaks at $r\sim 0.4$ and has a fast
rise at smaller $r$ and a slow decay for larger $r$. 
The mode of this pdf indicates that {\it kinematic ages are
systematically smaller than conventional spindown ages}.
This conclusion is in accord with our previous discussion,
where we invoked specific mechanisms to maximize the
likelihood in our velocity analysis and which also implied
smaller kinematic ages.
The results given in this 
section demonstrate this conclusion  more directly and also show
that it is a general trend in the population and not an artifact 
produced by a few extreme cases.  Also, the pdf for age is calculated
independently of any particular physical process, such
as braking mechanisms other than that due to magnetic dipole
radiation and torque decay. 

\section{CONSTRAINTS ON INDIVIDUAL PULSARS}\label{sec:pulsars}

In Appendix \ref{app:individuals} we discuss 
constraints and uncertainties for particular objects, especially those
with uncertain distances and large velocities.  Here we comment on
the possible birth sites and velocities of a few objects.

The Geminga pulsar (B0630+17) was discovered as a gamma-ray source
and eluded detection as a periodic source until 1992 
(Halpern \& Holt 1992).  Recent parallax and earlier proper-motion
observations (Caraveo 1993; Caraveo \etal 1996) yield fairly tight
determinations of its distance and proper motion.  Caraveo \etal 
discuss possible birth sites for this pulsar, including work
by Gehrels \& Chen (1993), Frisch (1993), Smith \etal (1994), and
Cunha \& Smith (1996).  The latter three authors suggest birth in
the event that has produced a supernova-like ring around 
$\lambda$Ori, a possible companion to the pulsar progenitor, now
requiring a radial velocity of about $-700$ \kms.  Our constraints
on the radial velocity (Table \ref{tab:components}),
$V_r \approx 0 \pm 178$ \kms suggests this association is
unlikely, essentially the same argument by Carveo \etal, who
state that the velocity would make an unlikely small angle with
the line of sight ($\sim 11^{\circ}$).

\section{A VELOCITY-MAGNETIC MOMENT CORRELATION?}\label{sec:vb}

Various kick mechanisms suggest the magnetic field of a young pulsar
is correlated with its acceleration.\footnote{According to the
``rocket theory'' of Harrison \& Tademaru (1975), pulsars with
off-axis magnetic dipoles are accelerated as they spin down.  For
rapidly spinning neutron stars, the final velocity is independent of
the magnetic moment since larger magnetic moments cause a larger force
that acts for a shorter time. The net velocity created by the
Harrison-Tademaru rocket depends on the birth period and offset of the
magnetic axis and one would not expect to find a correlation between a
pulsar's velocity today and its initial magnetic field.   However,
a relation between the offset and magnetic field strength would
introduce a correlation.} 
Horowitz and Piekarewicz (1996; see also Bisnovatyi-Kogan 1996) 
have discussed the role of parity violation for
neutrinos interacting with electrons and nucleons in a
strong magnetic field. Several interactions (neutrino scattering
from polarized nucleons, from polarized electrons, and in the presence
of a polarized background of electrons) can couple the momentum
flux of the neutrinos to the B field direction with a magnitude
proportional to B, potentially leading to the recoil of the
entire neutron star. A field of $\sim 10^{16}$ G is needed to
achieve the observed velocities. A more speculative suggestion of 
Kusenko \& Segr\`{e} (1996a) is that neutrino oscillations
are effected by the strength of the magnetic field. 
If the field is sufficiently strong ($\sim 10^{14}$ G according to the 
original proposal, $\sim 10^{16}$ G according to revised estimates
by Qian 1997), then interconversion of $\nu_e$ and $\nu_\tau$ between
the two different neutrinospheres allows some of the trapped electron
neutrinos to convert to the tau form and escape.
Neutron star recoil results from anisotropic emission of neutrinos
and the resultant momentum impulse. The resonant interconversion
depends on the function $\hat{k} \cdot \vec{B}$ and the {\it net}
momentum impulse could vanish for certain perfectly symmetric field
configurations (e.g. a strong, exactly toroidal field but not
a dipole field).
Kusenko \& Segr\`{e} (1996b) discussed an expected correlation in
their model, $v \propto B$ for long period pulsars.  Birkel and
Toldr\`{a} (1997) considered the correlation between transverse
velocity and the estimated magnetic field projected along the pulsar
spin axis for short period objects. The former find evidence in favor
of the hypothesis, the latter do not.

Some previous investigations have identified significant correlations
between perpendicular velocities and magnetic moment (e.g. Anderson
\& Lyne 1983; Cordes 1986, 1987; Bailes 1989) to
varying degree and others have not (Lorimer, Lyne \& Anderson 1995).
Using the derived velocities for our sample, 
we tested for a correlation between the 3D velocities of
Table~\ref{tab:components} and various combinations of $P$ and $\dot
P$ by cross correlating $\log V$ and $\log P\dot P^{\beta}$.
Figure~\ref{fig:corr} shows the correlation coefficient plotted
against $\beta$.  The correlation maximizes at $\beta = 0$ with a
correlation coefficient of 0.26, which can be achieved by chance at
the 7.8\% level.  This indicates that the 3D velocity is not
correlated with the magnetic moment ($\beta = 1$) or with the spindown
age ($\beta = -1$) or any other combination that involves a
significant contribution from $\dot P$.  Moreover, velocities are at
most weakly correlated with spin period.

Since there has been some disagreement over the significance of
the correlation in the past, we provide the following discussion.
The previously analyzed samples
included objects which have been spun up, i.e. millisecond
pulsars, and which have significantly lower space velocities 
and smaller magnetic fields than the
young, pulsars considered in this paper.  As shown in paper I, (Cordes
\& Chernoff 1997) the nominal rms 3D velocity for MSPs $\sim 80$ \kms
is a factor $\sim 5$ smaller than that of young pulsars.  
It seems clear that
the correlations between $V$ and $P\dot P$ previously identified were
effectively measuring this difference between the young and 
MSP
populations.  As argued in paper I, both the low magnetic fields and
space velocities of MSPs are related to the required binary
configuration needed in order for accretion driven spinup to occur. 
Accretion quenches or rearranges the magnetic field of the pulsar so
as to diminish its apparent dipolar magnetic field.  Binaries that are
not disrupted have low center of mass velocities. This economical
interpretation suggests that the apparent $V$-$P\dot P$ correlation
has nothing to do with the physics of core collapse that produces
neutron stars.  Bailes (1989) has suggested one scenario where a pair
of NSs are created from a binary.  The binary evolves in such a way to
produce one strong-field, high-velocity NS and one low-field,
low-velocity NS.  Our interpretation and our statistical results on
the kinematics of pulsars are consistent with this scenario but
do not require the particular assumptions of his model.

Although we find no evidence of correlation between $v$ and $B$, we do
not regard that as particularly strong evidence for or against the
role of neutrino-mediated parity violations or neutrino oscillations.
The magnetic field required for these schemes is $\sim 10^{14} -
10^{16}$ G, much larger than the inferred dipolar field today. Thus,
the original field must have decayed by $10^2 - 10^4$ and it is
unclear what relationship today's residual field bears to the
field that acted during the epoch of acceleration.

\section{PULSAR-SUPERNOVA REMNANT ASSOCIATIONS}\label{sec:psrsnr}

Frail, Goss \& Whiteoak (1994) deduce a mean transverse velocity $\sim
550$ \kms for 12 pulsars near supernova remnants, using the apparent
angular offset between pulsar and remnant centroid, a distance
estimate, and an age estimate, taken to be the conventional spindown
age.  One association in their sample (B$1757-24$ \& G$5.4-1.2$)
yields a transverse speed $\sim 1800$ \kms.  A few others were
excluded on the basis of their `extreme' velocities, ranging from 1600
to 3600 \kms, including an association between PSR B$1610-50$ and Kes
32 proposed by Caraveo (1993).

Our constraints on kinematic pulsar ages (\S\ref{sec:ages}) are based
on objects that are much older than the characteristic 20 kyr age of
those in the Frail \etal sample.  If there is no field decay, the most
likely braking index, $n \approx 4.5$, implies that the Frail \etal
velocities are biased too low by about a factor of two.  The mean
transverse speed in the Frail \etal sample would then be $\sim 1100$ \kms,
which is about 3 times higher than the mean transverse speed in our
best-fit model 2.1. If there is field decay on a timescale of $3$ Myr
and a braking index $n \approx 2.5$ (as in our best model), then the
Frail \etal sample will have ages that are slightly {\it larger} than
the conventional spindown times by a factor $3/2.5 = 1.2$, thus
reducing the mean, transverse velocity in Frail \etal by about 20\% to
$\sim 460$ \kms, which is still larger than the transverse mean ($\sim
331$ \kms) of our velocity distribution by 26\%.

In either case, the velocities calculated from pulsar-supernova remnant 
associations are larger than the mean pulsar distribution. 
This is intriguing and may
indicate that supernova remnants are produced preferentially by the
explosions that yield fast kicks.  It is clear that a statistical
analysis of the significance of the pulsar-supernova remnant
associations and a more complete analysis of the selection effects in
the pulsar surveys are needed. For example, some of the higher
velocity objects in the Frail \etal sample may be spurious
associations.  To gain some sense of the incompleteness of the pulsar
survey at high velocity, we can assume that the 12 associations used
by Frail \etal are real and estimate a correction to the pulsar birth
rate for the selection against high-velocity objects in pulsar surveys
as follows. Frail \etal estimate a median velocity $\sim 460$ \kms
that we correct to 380 \kms (using a braking index of 2.5).  From our
velocity pdf based on 47 pulsars with average spindown
age 2.95 Myr, we find
$P\{\vperp > 380\,\,{\rm km\,s^{-1} } \} = 0.22$.  Assuming identical
low velocity distributions, our sample is depleted of high velocity
pulsars by a factor $0.5/0.22 \sim 2.3$. For all velocities the sample
has been depleted by a factor $1/0.72 \sim 1.4$.  Thus, if the Frail
\etal estimates are correct, then about 40\% of pulsars are missed due
to selection against high-velocities.

\section{ASTROPHYSICAL CONSEQUENCES}\label{sec:astrophy}

In this section we discuss a few of the astrophysical implications
of the results we have reached.

\subsection{Evidence for Asymmetric Supernova Kicks}

The large peculiar velocities we have estimated for pulsars---and
their distribution--- are the net result of kicks from supernova
explosions, including rocket-like accelerations (Harrison \& Tademura
1975), and disrupted orbital motion in those cases where a neutron
star, or its progenitor, have become unbound from a binary or multiple
star system.

A considerable body of evidence supports the conclusion that supernova
explosions are, in general, asymmetric so as to impart  significant
velocities to neutron stars.  This evidence includes 
(1) statistical arguments based on population synthesis studies
which indicate that symmetric explosions yield too many binary
radio pulsars and too small a mean speed (Dewey \& Cordes 1987;
hereafter DC87);
(2) the existence of objects, like the Guitar Nebula pulsar
(B2224+65), which have velocities larger than that of any plausible
progenitor binary;
(3) the occurrence of geodetic precession in two NS-NS binaries,
which requires a misalignment of the spin and orbital angular
momenta through explosion asymmetry (Cordes \& Wasserman 1984; 
Weisberg \etal 1989; Cordes, Wasserman \& Blaskiewicz 1990;
Arzoumanian \etal 1996) suggestive of a kick $\sim 100-300$ \kms; 
(4) orbital precession in the pulsar J0045-7319, probably the result
of spin-orbit coupling (Lai \etal 1995; Kaspi \etal 1996).
Assuming aligned angular momenta in the pre-supernova state, Kaspi
\etal infer a probable minimum kick of 100 \kms to achieve the current
misalignment of spin and orbital momenta.

Quantitatively, except for the Guitar nebula pulsar, these facts each
call for kicks $\simgreat 100$ \kms. For example, DC87 concluded that
$\sim 90$ \kms kicks were needed to account for the then-observed
$\sim 100$ \kms pulsar velocity distribution and the low binary
frequency of pulsars. Current population synthesis models of neutron
star containing binaries predict birth rates of wide and close radio
pulsar binaries and of high mass X ray binaries in rough agreement
with observations when kicks are drawn from a Gaussian with dispersion
$\sim 450$ \kms (Portegies Zwart \& Verbunt 1996).  Without the kick,
far too many Be-pulsar binaries, Be- and high-mass X-ray binaries and
binary radio pulsars are found. It is important to note that the
minimum kick needed to remove the wide binaries is less than that
needed to explain the proper motions studied in this paper.  Also, the
only direct measurement of very fast $\sim 1000$ \kms objects is that
of the Guitar nebula pulsar. Indirect evidence comes from statistical
studies of proper motion like the current one and that of Lyne \&
Lorimer (1994). Less clear evidence comes from associations of pulsars
with some supernova remnants of known age (see \S\ref{sec:psrsnr}).

\subsection{High Velocities and Escape from the Galaxy}
\label{sec:fraction}

The current results extend previous results in a number of ways.  We
begin with the high velocity end of the pdf. The pulsar sample itself
includes 5 examples (11 \%) whose inferred 3D velocities exceed 1000
km s$^{-1}$.

Our model fitting allows us to estimate the fraction of objects that
have sufficient total velocity to escape the Galaxy.  The escape
velocity is, of course, not known with precision and it is a strong
function of location in the Galaxy.  Leonard \& Tremaine (1990) place
a lower bound on the local $V_e$ of 430 km s$^{-1}$.  Integrations of
the Paczynski potential also indicate that 500 km s$^{-1}$ is a
representative estimate for $V_e$, so we take this as a fiducial
value.

In Figure~\ref{fig:pdf_cdf} we show the pdfs for the peculiar speed
and the cumulative distribution functions (cdfs) of speed for three
models ($n_g.n_h = 1.1, 2.1$ \& 3.1).  The poorly-fitting,
single-component model (1.1) suggests that 50\% of the known, young
pulsars will escape the Galaxy.  The best model (2.1) and
slightly-worse model (3.1) yield a more modest $\sim 20$\%.  The cdf
for pulsar speeds indicates 5-25 \% above 1000 \kms, taking into
account uncertainties in our fitting, as indicated in
Figure~\ref{fig:pdf_cdf}.

In interpreting these results, it is necessary to reiterate that the
pulsar sample is known to be strongly biased by selection effects,
many of which arise from the fact that pulsar surveys are
signal-to-noise limited.  Many factors contribute to the net S/N for a
given pulsar, including period dependence of the underlying pulsar
beam shapes and luminosity.  More importantly, however, is the
velocity selection that occurs (Helfand \& Tademaru 1977; Cordes 1986;
Lorimer \& Lyne 1994): a greater number of fast pulsars will move out
of the detection volume centered on the Earth than will move into it.
Thus, pulsar surveys are biased against high velocity pulsars.  The
magnitude of this bias depends on details of specific pulsar surveys.
Our analysis does not incorporate the detection-related part of the
observer's selection function (see introduction) and it is not
possible to correct {\it post facto} the derived pdf.  We are
currently working on incorporating the detection-related terms to the
selection function (Chernoff \& Cordes 1997).  The velocity pdf
derived by Lyne \& Lorimer (1994) has an overall shape that is similar
to our double Gaussian model, but predicts a larger fraction of
pulsars that will escape the Galaxy, $>50\%$.

\subsection{Number of Separate Velocity Components}

The two component distribution, Model 2.1, we have found to be the
best fit in \S\ref{sec:odds}, may or may not correspond to two
physical subcomponents in the Galactic population of radio pulsars.
Our original assumption was that the sum-of-Gaussians form would
suffice to model any actual pdf, with no implication that the
individual Gaussians corresponded to distinct physical subpopulations.
However, the fact that the three component model has an odds ratio
inferior to the two component model and, in addition, that there is a
significant disparity between the individual velocity parameters
(${\sigma_V}_1 \sim 4{\sigma_V}_2$) is suggestive that two distinct
subpopulations may be involved. At the same time, we note that we have
explored only Gaussian models and that testing other parametric forms
will be necessary before drawing this conclusion.

To test for the presence of an additional low-velocity
population of pulsars, we explored solutions where two of the
components were varied near our best fit for the two-component model
(2.1) and a third additional component was imposed with dispersions in
the range 10 \kms $\le {\sigma_V}_3 \le$ 210 \kms.  By varying the
weights, $w_1, w_2$, dispersions ${\sigma_V}_1$, ${\sigma_V}_2$ while
holding the birth scale height (${h_z}_1$) fixed, we find $w_3 \le
0.02$, $0.05$, $0.1$ for ${\sigma_V}_3 \le 30$, $50$ and $70$ \kms,
respectively.  We therefore find no evidence in the data for a
low-velocity component that might provide a signature of symmetric
supernovae in isolated pulsars or in wide binaries.  There is only a
few percent allowed for pulsars with speeds less than 50 \kms,
consistent with the appearance of only two objects in our sample
(B1747$-$28 and B1929+10) whose 3D speeds could be, but need not be,
this small.

If there were no intrinsic kick then isolated supernovae would produce
low velocity neutron stars with velocities of order their progenitors'
speeds, $\sim 30$ \kms.  The fraction of NS that originate from
isolated (nonbinary) progenitors is not well known. In the models of
Portegies Zwart \& Verbunt (1996) composed of 40\% single stars, 60\%
binaries, with LL type kicks at birth (models SS and AK) approximately 34\%
of all supernova occur in isolated progenitors, 66\% in binaries. Since
our model implies that there are relatively few low velocity pulsars,
a natural interpretation is that all neutron stars receive a
substantial kick at birth. 

This conclusion is at variance with the recent proposal of Iben and
Tutukov (1996) that there are no intrinsic kicks during supernova collapse,
that instead, the pulsar velocity is produced entirely by the breakup of
binaries. To avoid overproducing low velocity pulsars, Iben and
Tutukov posit that pulsars can arise only  from stars in which mass
transfer or tidal interactions function to spin up the star prior to
collapse.  Comparison with the Portegies Zwart \& Verbunt results
shows that this assumption is not sufficient to zero out the low
velocity part of the pdf. Roughly half of the supernovae in binaries
involve mergers of stars prior to the explosion. We anticipate that
the merged objects will resemble single stars in their center of mass
motion but will contain substantial spin angular momentum.  If these
objects form neutron stars as the population synthesis models predict
and
if the neutron stars become visible pulsars but do not receive
birth kicks then they will contribute weight to the low velocity part
of the pdf. This is not what we observe. Thus, whether or not
single stars give rise to pulsars it appears that kicks $\simgreat 50$ \kms
are {\it generally required} for consistency with the pdf.

A supernova kick of $\sim 100$ \kms has generally been called for in
population synthesis models (DC87): otherwise too many neutron star
containing binaries are created. It is satisfying that the binary
abundance gives results consistent with the low velocity end of the
pulsar pdf. These abundances are an intrinsically less clear cut constraint
because of the many complexities of binary evolution including the
uncertainties implicit in the initial conditions.  On the other hand,
the small size of our current sample limits the precision of the direct
determination of the pdf at low velocity. Future proper motion
measurements should improve the situation.

Recent searches for old, accreting, isolated neutron stars have turned
up few prospects (see Wang 1996 for a review and Manning, Jeffries \&
Willmore 1996 for limits on the ROSAT Wide Field Camera survey).  Our
results imply that fewer than 10\% of neutron stars 
move with velocities less than 70 \kms (see above). This
differs from the estimates of $\sim 25-30$\% used by Madau \& Blaes
(1994) which were based on the young pulsar velocity distribution of
Narayan \& Ostriker (1990), diffusively heated by gravitational
interactions.  Since Bondi-Hoyle accretion scales as $v^{-3}$, the new
velocity distribution should play an important role in determining the
number of detected objects in X ray surveys and may help reconcile the
paucity of sources and the theoretical predictions. The accretion rate
onto the star depends also on the strength of the magnetic field and
the rotation rate, so our estimates of the braking index and torque
decay timescale are relevant too.

For supernovae in binaries that do not involve mergers (roughly
30\%) the resultant velocity distribution represents a convolution of
the orbital velocity distribution with the intrinsic kick velocity
distribution. The maximum velocity attainable with $90$ \kms kicks for
initial masses up to $20 \msun$ was estimated by DC87 to be about
$600$ \kms (by reaction [N1], with such a small separation that the
supernova shock completely destroys the companion) but this
constituted $<0.2$\% of all formed neutron stars. A more significant
$8$\% contribution of $\sim 500$ \kms was found in reaction [N3] in
which non-conservative evolution creates an unbound neutron star and
white dwarf.   These numbers are, of course, only examples based
on the input quantities used by DC87; especially important is their
attribution of final (pre-explosion) orbital periods after
the spiral-in processes that influence the N1 and N3 paths;  these
periods were only guessed at, implying that the typical 
500-600 \kms velocities estimated for these paths are not
hard numbers.

Iben and Tutukov (1996) argued that the
highest orbital velocity of neutron star are created when the
primary-descended neutron star in a binary is released as a
consequence of a second supernova explosion. They envision a
semi-detached binary consisting of a $16 \msun$ He core and an neutron
star with a $1.7 \rsun$ semi-major axis. The orbital velocity is $\sim
1300$ \kms and the first neutron star is ejected at $\sim 900$ \kms.
Given that the initial mass of the main sequence progenitor of the He
core is about $40 \msun$, close to the maximum mass for a star to
yield a neutron star, their example provides an upper bound to the
possible orbital velocity. Our result that 5 of the pulsars have
velocities in excess of $1000$ \kms then provides decisive evidence
that kicks of some sort are required.

In the future, once the observational selection effects are more fully
accounted for, it will be important to analyze the velocity pdf in
detail.  At the moment, we make a few general comments. An outstanding
issue is whether the two Gaussian description corresponds to a single
distribution with an extended high velocity tail or whether it more
closely corresponds to two separate populations of pulsars.
Our two component results imply low and high velocity
pulsars with comparable numerical weights (based on the crude estimate
of the selection against high velocity pulsars above).  If the
contribution from the kick and from the orbital disruption of binaries
were randomly combined (in the sense of two random velocities drawn
from individual Gaussians) then the resulting distribution would be
unimodal in speed with a dispersion that is the root mean square of
the individual dispersions. The current result is not unimodal.

This may be an indication that the two known velocity sources, orbital
motion and intrinsic supernova kick size, are correlated.  For
example, binaries in which super giants explode will generally be
larger than those in which He cores explode (since mass transfer has
occurred in the second but not the first case). If the kick size of
the former is smaller than the latter, then the pdf for velocity may
have two distinct components. In fact, supernovae will occur in a
variety of modified stellar types in binaries (He core, CO core,
Wolf-Rayet and super giant according to Portegies Zwart \& Verbunt
1996) and more complicated correlations with orbital motion can be
expected.  Another possibility is that the merged stars with no
orbital motion but large angular momentum may experience kicks of a
different characteristic size than regular or stripped stars.  While
all these possibilities are speculative, the accurate determination of
the pdf, whether or not it is unimodal, promises to be of value in
elucidating the complexities of binary evolution.

\subsection{Implications for core collapse}\label{sec:collapse}

An important aspect of this work is that we establish that a
significant fraction of the pulsar velocity distribution lies
above 1000 \kms. This should provide important constraints on
numerical simulations of supernova explosions.

In a supernova collapse the energy scales are $10^{53}$ ergs for
neutrino emission and neutron star binding energy, $10^{51}$ ergs for
mass motions plus optical display, and $1.4 \times 10^{49}$ ergs for a
1.4 solar mass neutron star moving at 1000 \kms. Thus, the energy
involved in NS motion is but a small fraction of the total.  Moreover,
the NS velocity is quite small relative to free fall velocities in the
vicinity of the core. It is interesting to note that the distance
traveled by the core moving at a typical speed of $1000$ \kms for an
interval corresponding to the delay before explosion (0.07-0.42 s for
SN1987a as estimated by Bethe 1993) is substantial compared to the
characteristic dimensions of the system. Bethe estimates three
important lengthscales: the neutrinosphere where the optical depth to
neutrinos becomes of order unity ($20$ km), the gain radius where
heating by the radiated neutrinos exceeds the gas' cooling ($\sim 140$
km) and the free nucleon radius where alpha particles are completely
dissociated into neutrons and protons ($\sim 210$ km). As the
protoneutron star forms, the accretion shock stands off at $\sim 300$
km and begins to accelerate outward once the ram pressure of the
accreting material diminishes sufficiently. The neutrino heating and
convection behind the shock are essential for providing the pressure
that launches the shock. The core's motion is slow and the material
between the neutron star and the shock is subsonic, so whatever
acceleration mechanism functions, it should make little difference for
estimating the state of the normal matter within the shock.

We begin by describing the limits to the generation of core motion.
If the kick $\Delta v_{ns}$ is ultimately due to the momentum impulse associated with
anisotropic neutrino emission, then a firm limit may be expressed in
terms of the total binding energy of the neutron star, e.g.
\begin{equation}
\Delta v_{ns} < \epsilon_\nu {\Delta E \over M_{ns} c^2} c
\end{equation}
where $\epsilon_\nu$ is the anisotropy in the neutrino momentum and
$\Delta E$ is the total energy emitted, which must be less than the
binding energy of the final neutron star $\approx 0.15 M_{ns}c^2$.  To
give kicks of $1500$ \kms, we require that $ \epsilon_\nu > 0.03$.
From numerical simulations, Janka \& M\"uller estimate that the energy
that is available is significantly less than the binding energy --
more closely the accretion of the mantle during the convective period:
$\Delta E \approx 0.015 M_{ns}c^2$. Then a
substantially larger asymmetry $ \epsilon_\nu > 0.3$ is required to give
$1500$ \kms. This degree of asymmetry is inconsistent with
simulations.

If the kick is due to the momentum recoil from either accreted or
expelled matter, the momentum transfer is of
order the escape velocity near the gain radius times the mass
contained between the gain region and the accretion shock front times an
asymmetry factor, e.g.
\begin{equation}
\Delta v_{ns} < \epsilon_{ej} {\Delta M \over M_{ns}} v_{ej}
\end{equation}
where $\epsilon_{ej}$ is the anisotropy in the momentum carried off,
$\Delta M$ is the mass and $v_{ej}$ is the escape velocity. From
numerical simulations, Janka \&
M\"uller estimate from the most extreme cases that $\Delta M/M_{ns} <
0.2$, $v_{ej} < 0.1 c$ and $\epsilon_{ej} < 0.08$ and infer $\Delta
v_{ns} < 500$ \kms. Once again, a substantially larger asymmetry
$\epsilon_{ej} > 0.24$ would be required to give $1500$ \kms.

These general arguments are not without loopholes, of course. They
depend on the validity of the existing numerical simulations to set
reasonable limits for $\epsilon_\nu$ or $\epsilon_{ej}$.  In all
calculations to date the neutron star is fixed at the center of
symmetry. As mentioned above, because the fastest neutron star motions
are subsonic these approximations should be reasonable for gauging the
size of purely hydrodynamic
quantities like $\Delta M$, $\Delta E$, $v_{ej}$ and
$\epsilon_{ej}$ that appear above. The fact that neutrino
transport is treated with limited accuracy and that current numerical
modeling is not done in full 3D are much more difficult issues to
evaluate. 

As many others have discussed, it would appear that the likely
physical sources for a kick are (1) the large neutrino flux
beyond the neutrinosphere and
(2) the outer envelope of infalling material which is not in sonic
communication with the matter surrounding the core.  Goldreich, Lai \&
Sahrling (1997) have suggested that core g modes in presupernova stars
may be overstable. As the matter collapses, irregularities in the
infalling envelope may grow to large amplitude and ultimately impart a
kick by gravitational torques.  Another possibility is envisioned by
Burrows and Hayes (1996) in which the inhomogeneities of the infalling
matter channel the supernova blast into a jet-like outflow that causes
the core to recoil. 

It is important to point out, also, that the lack of correlation
between the velocity and magnetic field in today's pulsars
(\S\ref{sec:vb}) does not necessarily rule out a neutrino-magnetic
field interaction ultimately responsible for the recoil. If field
decay has occurred, as suggested by our best fitting models with
torque decay, then the correlation between magnetic field and velocity
could be weakened.

\section{SEARCHING FOR THE HIGHEST-VELOCITY PULSARS}\label{sec:searching}

It is almost certain that the velocity pdf we have derived underestimates
the fraction of pulsars with $V \simgreat 10^3$ \kms.   Searching for
high velocity pulsars at high galactic latitudes is a clear, though
nontrivial,  way to improve the situation.  The difficulty arises
because 
(1) such pulsars will be faint owing to their presumably large distances; 
(2) proper motion measurements will require VLBI techniques; 
    given pulsar faintness, the VLBA will have to be augmented
    by the large-aperture Arecibo Telescope and Green Bank Telescope; 
(3) high-velocity pulsars will be above the thickest free electron 
    layer in the Galaxy and therefore their distances cannot be
    estimated from the TC (or any other) model for the electron density.
Difficulties (1) and (2) can be surmounted through suitable allotment
of telescope time.   Estimating pulsar distances is much more 
difficult to tackle, for inverting dispersion measures will never be
possible for pulsars with $\vert z \vert \simgreat 1$ kpc.  Parallax
observations with VLBI or pulse-timing techniques are also unlikely.
The only possible recourse is development of an alternative distance
scale.   We are investigating the possibility that radio pulsars
have beam components whose luminosities are determined by
$P$ and $\dot P$ and no other parameters.  This `standard candle' approach
also requires that beam shapes and angular diameters also be predictable,
in which case orientation angles between the pulsar spin and magnetic axes
and the observer's direction can be determined, along with the predicted
flux and, hence, distance.   Work on this beam modeling will be reported 
elsewhere.   Should beam modeling fail due to (e.g.) pulsar
`weather' effects, such as severe distortions of magnetic fields from dipolar
forms by accretion or other history dependent activity, then 
analysis of high latitude pulsars must rely on statistics of large numbers
of objects rather than velocity estimates of individual objects.

\section{SUMMARY AND CONCLUSIONS}\label{sec:summary}

In this paper we have used the data on 47 pulsars to infer their
velocity and z-altitude distributions at birth.  Our likelihood
method allows estimation of radial as well as transverse velocities,
contingent on estimates for the ages and birth z's of individual 
objects.  We emphasize that we have not corrected the results for
the strong selection effects in the original pulsar surveys that
have provided the sample of objects.  The strongest effect on 
our conclusions is that we have underestimated the fraction of
pulsars with large velocities, e.g. those with $V > 1000$ \kms.
We are currently developing methods, similar to those in this
paper and in Paper I, that take selection effects into account. 

With these provisos, we have found that:
\begin{enumerate}
\item All pulsars in the sample are consistent with birth within
0.3 kpc of the galactic plane; any apparent motion of pulsars
toward the galactic plane is easily explainable as a combination
of projection effects associated with pulsars that are relatively
nearby, a conclusion made long ago, on a smaller sample,
by Helfand \& Tademaru (1977).
\item The best fit model we have found consists of two 
Gaussian components in velocity, with dispersions 
$\sim 175$ and 700 \kms, combined with a birth scale height
of 0.13 kpc;  we have not investigated mathematical forms other
than the multicomponent Gausssians. 
\item With our fit, we predict that $\sim 20\%$ of pulsars like those
in our sample will escape the Galaxy;  a rough estimate indicates that
this fraction may be underestimated by a factor $\sim 2$ due to
selection effects.
\item A braking index $n\sim2.5$ is favored if the pulsar torque
constant decays on a time scale $\sim 3$ Myr.  If no torque decay
occurs, then most braking indices are in the range 
$3.5$-$6$, though such a model yields a slightly inferior
likelihood than does the model with torque decay.
\item The constraints on braking index and torque decay time arise
from the role that the kinematic age plays in our estimates of
radial velocities.  The most general statement is that pulsar ages
are typically a factor $\sim 2$ smaller than the conventional
spindown ages calculated as $P/2\dot P$, which assumes birth
at a period much smaller than the present day period and a braking 
index of 3, with no torque decay.
\item We find no significant correlation between velocity and
magnetic field estimates.  Reconciling this result with previous
studies is simple:  previous work included strong field and weak
field objects that had been spun up by accretion.  We found
in Paper I that millisecond pulsars are generally a low velocity 
population.    The correlation found in previous work appears
to signify only the different evolutionary paths that have led
to strong field objects and spun-up, low field objects.   
\item Pulsar velocities require some sort of rocket mechanism
that augments velocities that would obtain under the condition
of symmetric supernovae in binary and nonbinary progenitors.
No strong discrimination exists in our analysis for proposed
rocket effects that take place during or shortly after supernova
explosions.  However, the estimated velocities require
a more efficient acceleration mechanism than has been calculated
thus far.
\item We suggest that the best means for pushing the limits and testing
models is to search for, and measure the proper motions of,
pulsars at high galactic latitudes.
\end{enumerate}

Acknowledgements:

We thank Z. Arzoumanian, D. Lorimer, T. Loredo \& I. Wasserman for
useful discussions.  This work was supported by NSF Grants
AST-9530397, AST-9528394 and NASA Grant 5-2851.  This research was
conducted using the resources of the Cornell Theory Center, which
receives major funding from the National Science Foundation (NSF) and
New York State, with additional support from the Advanced Research
Projects (ARPA), the National Center for Research Resources at the
National Institutes of Health (NIH), IBM Corporation and other members
of the center's Corporate Partnership Program.  It was also supported
by the National Astronomy and Ionosphere Center, which operates the
Arecibo Observatory under a cooperative agreement with the National
Science Foundation.

\eject
\appendix
\centerline{\bf APPENDIX}
\section{Individual Objects}\label{app:individuals}

Distances, if misestimated by the TC model, are more likely to be 
overestimated than underestimated because HII regions or 
other enhancements of the  free-electron density will perturb the 
dispersion measure (DM) upwards.
In some cases, the distance estimate can be grossly overestimated.

\begin{itemize}
\item[] {\bf B0531+21:}
		  The Crab pulsar clearly was born near its present
                  $z$ distance from the galactic plane ($\sim -0.2$ kpc).
		  However, its progenitor star may have originated
		  much nearer the plane. Gott \etal (1970) argued
 		  that the Crab's progenitor and the 3.75 s pulsar
		  B0525+21 were unbound from the explosion that 
		  produced the latter pulsar;  the point of origin
		  was the Gem I OB association.
\item[] {\bf B0540+23:} A modest $\vr < 500$ \kms results if the pulsar
                  was born at $\zb\sim -0.3$ kpc.
\item[] {\bf B0630+17:} The Geminga pulsar is one of the few with a parallax
		  as well as proper motion measurement.  Its speed is modest
		  and the likely range of radial velocities renders
		  unlikely any association with the star $\lambda$Ori,
		  which would require a radial velocity of about
		  $-700$ \kms (Caraveo \etal 1996).   A histogram of
		  Monte Carlo values of radial velocity 
		  (cf. \S\ref{sec:veluno}) shows only 0.3\% of counts
		  have radial velocity magnitudes as large as
		  700 km s$^{-1}$.
\item[] {\bf B0736$-$40:} 
		  The TC model yields only a lower bound on the 
		  distance (11.0 kpc) because its DM is too large
		  for its latitude.   Using this lower bound, the 
		  required radial velocity is 
		  $\vert \vr \vert > 10^4$ \kms.   The distance is
	 	  likely to be much smaller because
		  an HII region may account for much of DM.  Frail (1990)
	          estimates the distance to be only 0.4 kpc based
		  on an analysis of HII regions along the line of sight.  
\item[] {\bf B1508+55:}
		  The catalog uncertainty in distance is 60\%.  The transverse
		  speed exceeds $\vr$ by a factor of 7.
\item[] {\bf B1642$-$03:}
		  Prentice \& ter Haar (1969) first proposed that the
		  DM to this pulsar is contributed to significantly by
                  an HII region.  Its distance is then 0.1-0.2 kpc rather
                  than the 1.6 kpc from the TC model.   We have used this
 		lower distance range.
\item[] {\bf B1706$-$16:}
		  Frail (1989) argues that an HII region can account for
 		  some of the DM for this pulsar, which might place the
		  pulsar significantly nearer ($\sim 0.1$ kpc) than
		  its nominal distance, 1.27 kpc.  This possibility is less
		  certain than that for B1642$-$03, so we have used the 
		  larger distance.  
\item[] {\bf B1933+16:} 
		  No auxiliary data contradict the adopted distance
		  or the large transverse speed.   The radial velocity
		  need not be large if the birth altitude $\sim 0.15$ kpc.
\item[] {\bf B1953+50:} 
		  The radial velocity exceeds the transverse speed by
		  a factor $\sim 3.5$.   No auxiliary data contradict the
		  assumed distance.   To reduce $\vert \vr \vert$ to
		  less than $10^3$ or 500 \kms, the distance would have 
		  to be reduced $<1.1$ kpc or $< 0.54$ kpc. 
		  Alternatively, for a distance equal to the midpoint,
		  $0.5(D_L+ D_U)=1.84$ kpc,  the age must be  
		  reduced by a factor $\sim 4-6$ to reduce the 
		  radial velocity magnitude to 1000 or 500 \kms.
\item[] {\bf B2154+40:} 
		  The radial velocity magnitude is about twice the 
		  transverse speed.   No auxiliary data exist to contradict
		  the derived distance and velocities. 
\item[] {\bf B2021+51:}
		  The parallax from VLBI measurements
		  (Campbell \etal 1996) yields a distance consistent with
		  that from the TC model.  The spindown age agrees with
		  the kinematic age to better than a factor of two.  
\item[] {\bf B2224+65:} 
		  The Guitar-Nebula pulsar is the largest, convincing
		  pulsar velocity in excess of 800 \kms 
		  (Cordes, Romani \& Lundgren 1993). At the TC distance,
		  the implied radial velocity and birth $z$ are both modest.
		  Evidently, most of the pulsar's speed is transverse to the 
	          line of sight and parallel to the galactic plane.  It will
		  escape the Galaxy.
\end{itemize}

\begin{table}
\caption{\hfil Pulsar Proper Motion Data}
\label{tab:psrdata}
\begin{center}
\begin{tabular}{llrrrrrrrrrrr}
\tableline
\\ 
\multicolumn{2}{c}{Name} & $\ell$ & b & $\mu_{\alpha}$ & $\mu_{\delta}$ & $\log\tau_S$ & $ D_L$ & $ D_U$ & Ref \\
(J2000)&(B1950) &  (deg) &(deg)& ($\frac{mas}{yr}$) & ($\frac{mas}{yr}$) & (yr)        & (kpc)  & (kpc)  \\
\\ 
\tableline
\tableline 
$0139+5814$ & $ 0136+57$ & $  129.2$ & $   -4.0$ & $  -11.0\pm    5.0$ & $  -19.0\pm     5.0$ &     5.6 &     2.22 &     3.76 &    1 \\  
$0332+5434$ & $ 0329+54$ & $  145.0$ & $   -1.2$ & $   17.0\pm    1.0$ & $  -13.0\pm     1.0$ &     6.7 &     1.10 &     1.86 &    1 \\  
$0358+5413$ & $ 0355+54$ & $  148.2$ & $    0.8$ & $   15.0\pm   10.0$ & $   -6.0\pm    10.0$ &     5.8 &     1.59 &     2.69 &    2 \\  
$0454+5543$ & $ 0450+55$ & $  152.6$ & $    7.5$ & $   52.0\pm    6.0$ & $  -17.0\pm     2.0$ &     6.4 &     0.61 &     1.03 &    1 \\  
$0502+4654$ & $ 0458+46$ & $  160.4$ & $    3.1$ & $   -8.0\pm    3.0$ & $    8.0\pm     5.0$ &     6.3 &     1.37 &     2.31 &    1 \\  
$0528+2200$ & $ 0525+21$ & $ -176.1$ & $   -6.9$ & $  -20.0\pm   19.0$ & $    7.0\pm     9.0$ &     6.2 &     1.75 &     2.95 &    1 \\  
$0534+2200$ & $ 0531+21$ & $ -175.4$ & $   -5.8$ & $  -16.0\pm   11.0$ & $   -2.0\pm     8.0$ &     3.1 &     1.54 &     2.60 &    3 \\  
$0543+2329$ & $ 0540+23$ & $ -175.6$ & $   -3.3$ & $   19.0\pm    7.0$ & $   12.0\pm     8.0$ &     5.4 &     2.72 &     4.60 &    1 \\  
$0614+2229$ & $ 0611+22$ & $ -171.2$ & $    2.4$ & $   -4.0\pm    5.0$ & $   -3.0\pm     7.0$ &     5.0 &     3.63 &     6.14 &    1 \\  
$0629+2415$ & $ 0626+24$ & $ -171.2$ & $    6.2$ & $   -7.0\pm   12.0$ & $    2.0\pm    12.0$ &     6.6 &     3.59 &     6.07 &    1 \\  
$0630-2834$ & $ 0628-28$ & $ -123.0$ & $  -16.8$ & $   -5.0\pm   18.0$ & $  -17.0\pm    26.0$ &     6.4 &     1.65 &     2.78 &    2 \\  
$0633+1746$ & $ 0630+17$ & $ -164.9$ & $    4.3$ & $  138.0\pm    4.0$ & $   97.0\pm     4.0$ &     5.5 &     0.12 &     0.22 &    4 \\  
$0653+8051$ & $ 0643+80$ & $  133.2$ & $   26.8$ & $   19.0\pm    3.0$ & $   -1.0\pm     3.0$ &     6.7 &     2.32 &     3.93 &    1 \\  
$0659+1414$ & $ 0656+14$ & $ -158.9$ & $    8.3$ & $   64.0\pm   11.0$ & $  -28.0\pm     7.0$ &     5.0 &     0.58 &     0.99 &    5 \\  
$0738-4042$ & $ 0736-40$ & $ -105.8$ & $   -9.2$ & $  -80.0\pm   13.0$ & $   -3.0\pm    43.0$ &     6.6 &     0.40 &     0.80 &    2 \\  
$0742-2822$ & $ 0740-28$ & $ -116.2$ & $   -2.4$ & $  -29.0\pm    0.9$ & $   -0.1\pm     0.4$ &     5.2 &     1.40 &     7.70 &    6 \\  
$0820-1350$ & $ 0818-13$ & $ -124.1$ & $   12.6$ & $    9.0\pm   10.0$ & $  -31.0\pm     7.0$ &     7.0 &     1.89 &     3.20 &    1 \\  
$0826+2637$ & $ 0823+26$ & $ -163.0$ & $   31.7$ & $   61.0\pm    3.0$ & $  -90.0\pm     2.0$ &     6.7 &     0.29 &     0.49 &    3 \\  
$0835-4510$ & $ 0833-45$ & $  -96.4$ & $   -2.8$ & $  -48.0\pm    2.0$ & $   34.9\pm     1.0$ &     4.0 &     0.39 &     0.65 &    6 \\  
$0837+0610$ & $ 0834+06$ & $ -140.3$ & $   26.3$ & $    2.0\pm    5.0$ & $   51.0\pm     3.0$ &     6.5 &     0.55 &     0.94 &    3 \\  
\tableline
\end{tabular}
\end{center}
\end{table}

\addtocounter{table}{-1}
\begin{table}
\caption{\hfil Pulsar Proper Motion Data (continued)}
\begin{center}
\begin{tabular}{llrrrrrrrrrrr}
\tableline
\\ 
\multicolumn{2}{c}{Name} & $\ell$ & b & $\mu_{\alpha}$ & $\mu_{\delta}$ & $\log\tau_S$ & $ D_L$ & $ D_U$ & Ref \\
(J2000)&(B1950) &  (deg) &(deg)& ($\frac{mas}{yr}$) & ($\frac{mas}{yr}$) & (yr)        & (kpc)  & (kpc)  \\
\\ 
\tableline
\tableline 
$0908-1739$ & $ 0906-17$ & $ -113.9$ & $   19.9$ & $   27.0\pm   11.0$ & $  -40.0\pm    11.0$ &     7.0 &     0.48 &     0.82 &    1 \\  
$0946+0951$ & $ 0943+10$ & $ -134.6$ & $   43.1$ & $  -38.0\pm   19.0$ & $  -21.0\pm    12.0$ &     6.7 &     0.75 &     1.27 &    3 \\  
$1136+1551$ & $ 1133+16$ & $ -118.1$ & $   69.2$ & $ -102.0\pm    5.0$ & $  357.0\pm     3.0$ &     6.7 &     0.21 &     0.35 &    3 \\  
$1453-6413$ & $ 1449-64$ & $  -44.3$ & $   -4.4$ & $  -16.4\pm    1.1$ & $  -21.3\pm     0.8$ &     6.0 &     2.00 &     5.00 &    6 \\  
$1509+5531$ & $ 1508+55$ & $   91.3$ & $   52.3$ & $  -73.0\pm    4.0$ & $  -68.0\pm     3.0$ &     6.4 &     1.49 &     2.51 &    3 \\  
$1559-4438$ & $ 1556-44$ & $  -25.5$ & $    6.4$ & $   11.0\pm   17.0$ & $   20.0\pm    50.0$ &     6.6 &     1.50 &     2.50 &    2 \\  
$1645-0317$ & $ 1642-03$ & $   14.1$ & $   26.1$ & $   41.0\pm   17.0$ & $  -25.0\pm    11.0$ &     6.5 &     0.10 &     0.50 &    3 \\  
$1709-1640$ & $ 1706-16$ & $    5.8$ & $   13.7$ & $   75.0\pm   20.0$ & $  147.0\pm    50.0$ &     6.2 &     0.98 &     1.65 &    2 \\  
$1752-2806$ & $ 1749-28$ & $    1.5$ & $   -1.0$ & $   -5.0\pm   17.0$ & $    8.0\pm    15.0$ &     6.0 &     1.18 &     1.99 &    2 \\  
$1820-0427$ & $ 1818-04$ & $   25.5$ & $    4.7$ & $    3.0\pm    3.0$ & $   27.0\pm     3.0$ &     6.2 &     1.61 &     2.73 &    3 \\  
$1825-0935$ & $ 1822-09$ & $   21.4$ & $    1.3$ & $   10.0\pm   19.0$ & $  -23.0\pm    19.0$ &     5.4 &     0.78 &     1.31 &    2 \\  
$1844+1454$ & $ 1842+14$ & $   45.6$ & $    8.1$ & $   -9.0\pm   10.0$ & $   45.0\pm     6.0$ &     6.5 &     1.70 &     2.87 &    1 \\  
$1913-0440$ & $ 1911-04$ & $   31.3$ & $   -7.1$ & $    7.0\pm   13.0$ & $   -5.0\pm     9.0$ &     6.5 &     2.48 &     4.19 &    1 \\  
$1919+0021$ & $ 1917+00$ & $   36.5$ & $   -6.2$ & $   -2.0\pm   30.0$ & $   -1.0\pm    10.0$ &     6.4 &     2.56 &     4.33 &    1 \\  
$1932+1059$ & $ 1929+10$ & $   47.4$ & $   -3.9$ & $   79.0\pm    6.0$ & $   39.0\pm     4.0$ &     6.5 &     0.13 &     0.22 &    3 \\  
$1935+1616$ & $ 1933+16$ & $   52.4$ & $   -2.1$ & $    2.0\pm    3.0$ & $  -25.0\pm     5.0$ &     6.0 &     6.11 &    10.32 &    7 \\  
$1948+3540$ & $ 1946+35$ & $   70.7$ & $    5.0$ & $   -9.0\pm    7.0$ & $   -4.0\pm     8.0$ &     6.2 &     6.05 &    10.22 &    1 \\  
$1955+5059$ & $ 1953+50$ & $   84.8$ & $   11.6$ & $  -23.0\pm    5.0$ & $   54.0\pm     5.0$ &     6.8 &     1.37 &     2.31 &    1 \\  
$2022+2854$ & $ 2020+28$ & $   68.9$ & $   -4.7$ & $   -9.0\pm    3.0$ & $  -13.0\pm     2.0$ &     6.5 &     1.00 &     1.69 &    3 \\  
$2022+5154$ & $ 2021+51$ & $   87.9$ & $    8.4$ & $   -8.1\pm    0.2$ & $   13.4\pm     0.2$ &     6.4 &     0.94 &     1.59 &    8 \\  
\tableline
\end{tabular}
\end{center}
\end{table}

\addtocounter{table}{-1}
\begin{table}
\caption{\hfil Pulsar Proper Motion Data (continued)}
\begin{center}
\begin{tabular}{llrrrrrrrrrrr}
\tableline
\\ 
\multicolumn{2}{c}{Name} & $\ell$ & b & $\mu_{\alpha}$ & $\mu_{\delta}$ & $\log\tau_S$ & $ D_L$ & $ D_U$ & Ref \\
(J2000)&(B1950) &  (deg) &(deg)& ($\frac{mas}{yr}$) & ($\frac{mas}{yr}$) & (yr)        & (kpc)  & (kpc)  \\
\\ 
\tableline
\tableline 
$2048-1616$ & $ 2045-16$ & $   30.5$ & $  -33.1$ & $   85.0\pm   16.0$ & $  -43.0\pm    17.0$ &     6.5 &     0.49 &     0.83 &    2 \\  
$2055+3630$ & $ 2053+36$ & $   79.1$ & $   -5.6$ & $   -3.0\pm    7.0$ & $    3.0\pm     3.0$ &     7.0 &     4.28 &     7.23 &    1 \\  
$2113+2754$ & $ 2110+27$ & $   75.0$ & $  -14.0$ & $  -23.0\pm    2.0$ & $  -54.0\pm     3.0$ &     6.9 &     1.05 &     1.78 &    1 \\  
$2157+4017$ & $ 2154+40$ & $   90.5$ & $  -11.3$ & $   18.0\pm    1.0$ & $   -3.0\pm     1.0$ &     6.8 &     4.25 &     7.19 &    1 \\  
$2219+4754$ & $ 2217+47$ & $   98.4$ & $   -7.6$ & $  -12.0\pm    8.0$ & $  -30.0\pm     6.0$ &     6.5 &     1.89 &     3.19 &    7 \\  
$2225+6535$ & $ 2224+65$ & $  108.6$ & $    6.8$ & $  144.0\pm    3.0$ & $  112.0\pm     3.0$ &     6.0 &     1.54 &     2.60 &    1 \\  
$2354+6155$ & $ 2351+61$ & $  116.2$ & $   -0.2$ & $   22.0\pm    3.0$ & $    6.0\pm     2.0$ &     6.0 &     2.53 &     4.28 &    1 \\  
\tableline
\end{tabular}

References:
1. Harrison \etal (1992)  
2. Fomalont \etal (1992)  
3. Lyne, Anderson \& Salter (1982)
4. Caraveo \etal (1996)  
5. Thompson \& Cordova (1994) 
6. Bailes \etal (1990)  
7. Downs \& Reichley (1983 
8. Campbell \etal (1996)  
\end{center}
\end{table}

\begin{table}
\caption{\hfil Parameter Search Ranges}
\label{tab:ranges}
\begin{center}
\begin{tabular}{cllll}
\tableline
\\
parameter & units & min & max & range \\
\\
\tableline
\tableline
\\
$w_1$ & --- & 0  & 1  &  $\Delta w_1 = 1$ \\
\\
$w_2$ & --- & 0  & 1  &  $\Delta w_2 = 1$ \\
\\
${h_z}_1$  & kpc  & 0.01 & 0.51  & $\Delta {h_z}_1 = 0.5$ \\
\\
${h_z}_2$  & kpc  & 0.01 & 0.51  & $\Delta {h_z}_2 = 0.5$ \\
\\
${\sigma_V}_1$ & km s$^{-1}$ & 25 & 2025 & $\Delta{\sigma_V}_1 = 2000$ \\
\\
${\sigma_V}_2$ & km s$^{-1}$ & 200 & 2000 & $\Delta{\sigma_V}_2 = 1800$ \\
\\
${\sigma_V}_3$ & km s$^{-1}$ & 100 & 400 & $\Delta{\sigma_V}_3 = 300$ \\
\\
\tableline
\end{tabular}
\end{center}
\end{table}

\begin{table}
\caption{3D Velocity PDF Models}
\label{tab:compare}
\begin{center}
\begin{tabular}{rllcccrcccl}
\tableline
Model  & $w_1$ & $w_2$ & ${h_z}_1$ & ${h_z}_2$ & ${\sigma_V}_1$ &
\multicolumn{1}{c}{${\sigma_V}_2$} & ${\sigma_V}_3$ & 
$N_{parms}$ & $\log {\cl L}$ & Odds\\
$n_g.n_h$    &       &       &  (kpc)    & (kpc)     & (km s$^{-1}$)  &
\multicolumn{1}{c}{(km s$^{-1}$)}  & (km s$^{-1}$)  & \\
\tableline
\tableline
\\
\multicolumn{4}{l} {$\taus < 10$ Myr: $N_{psr} =$ 47}\\
\cline{1-4}
\\
1.1 & 1 & ---  & 0.13 & --- & 300 & ---  & ---  & 2 & $-$263.50& 1\\
2.1 & 0.8 & 0.2 & 0.13 & --- & 175 & 700 & ---  & 4 & $-$255.92& $10^{6.3}$ \\
2.2& 0.82 & 0.18 & 0.13 & 0.10&  175 &  750 & ---  & 5 & $-$255.89& $10^{5.5}$ \\
3.1 & 0.3 & 0.15 & 0.13 & --- & 156 &  750 & 200  & 6 & $-$255.90& $10^{5.4}$ \\
\tableline
\\
\multicolumn{4}{l} {$\taus < 5$ Myr: $N_{psr} =$ 38}\\
\cline{1-4}
\\
1.1 & 1 & ---  & 0.13 & --- & 300 & ---  & ---  & 2 & $-$208.10& 1 \\
2.1 & 0.87 & 0.13 & 0.13 & --- & 175 & 700 & ---  & 4 & $-$201.94& $10^{5.1}$ \\
\tableline
\\
\multicolumn{4}{l} {$\taus < 1$ Myr: $N_{psr} =$ 12}\\
\cline{1-4}
\\
1.1 & 1 & ---  & 0.13 & --- & 200 & ---  & ---  & 2 & $-$55.99  & 1 \\
2.1 & 0.8  & 0.2  & 0.13 & --- & 175 & 400 & ---  & 4 & $-$55.95  & $0.36$ \\
\tableline

\end{tabular}
\end{center}
\end{table}

\begin{table}
\caption{\hfil Best-fit Parameters for Model 2.1}
\label{tab:bestfit}
\begin{center}
\begin{tabular}{cll}
\tableline
\\
parameter & value & units \\
\\
\tableline
\tableline
\\
$w_1$            & $0.83^{+0.03}_{-0.13}$   & --- \\
\\
${h_z}_1$        & $0.13^{+0.02}_{-0.03}$   & kpc  \\
\\
${\sigma_V}_1$   & $175^{+20}_{-26}$        & km s$^{-1}$ \\
\\
${\sigma_V}_2$   & $700^{+224}_{-148}$             & km s$^{-1}$ \\
\\
\tableline
\end{tabular}
\end{center}
\end{table}

\begin{table}
\caption{\hfil Pulsar Velocity Components}
\label{tab:components}
\begin{center}
\begin{tabular}{lrrrrr}
\tableline
\\ 
Name &  $ V\hfil$ &  $ V_r \hfil$ &  $ V_{\perp_1}\hfil$ & $ V_{\perp_2}\hfil$ & $ z_0 \hfil$ \\ 
\cline{2-5}
     & \multicolumn{4}{c}  {$\displaystyle {\rm (km \, s^{-1})} $}  & $(kpc) \hfil$  \\ 
\\ 
\tableline
\tableline 
$  0136+57$ &  $    349 \pm      148$ & $    -14 \pm      249$ & $   -139 \pm       53$ & $   -237 \pm         55$ & $   -0.10 \pm   0.032$  \\  
$  0329+54$ &  $    229 \pm      111$ & $    -24 \pm      203$ & $    119 \pm       19$ & $    -91 \pm         15$ & $    0.01 \pm   0.057$  \\  
$  0355+54$ &  $    239 \pm      141$ & $      5 \pm      216$ & $    139 \pm       69$ & $    -46 \pm         63$ & $    0.00 \pm   0.078$  \\  
$  0450+55$ &  $    263 \pm       74$ & $    -75 \pm      164$ & $    191 \pm       32$ & $    -65 \pm         11$ & $   -0.11 \pm   0.059$  \\  
$  0458+46$ &  $    193 \pm       96$ & $     10 \pm      183$ & $    -61 \pm       24$ & $     84 \pm         39$ & $    0.11 \pm   0.054$  \\  
$  0525+21$ &  $    290 \pm      130$ & $      3 \pm      204$ & $   -209 \pm       99$ & $     38 \pm         68$ & $   -0.04 \pm   0.163$  \\  
$  0531+21$ &  $    221 \pm      114$ & $     -2 \pm      194$ & $   -119 \pm       75$ & $    -13 \pm         65$ & $   -0.20 \pm   0.031$  \\  
$  0540+23$ &  $    341 \pm      149$ & $      3 \pm      249$ & $    215 \pm       73$ & $    133 \pm         84$ & $   -0.25 \pm   0.036$  \\  
$  0611+22$ &  $    223 \pm      120$ & $    -16 \pm      198$ & $    -58 \pm       86$ & $    -36 \pm        114$ & $    0.20 \pm   0.029$  \\  
$  0626+24$ &  $    312 \pm      242$ & $     71 \pm      332$ & $     92 \pm       86$ & $     78 \pm        136$ & $    0.04 \pm   0.168$  \\  
$  0628-28$ &  $    349 \pm      188$ & $    109 \pm      289$ & $   -141 \pm       88$ & $   -113 \pm        147$ & $   -0.05 \pm   0.139$  \\  
$  0630+17$ &  $    206 \pm       92$ & $      0 \pm      178$ & $    113 \pm       15$ & $     80 \pm         11$ & $   -0.03 \pm   0.006$  \\  
$  0643+80$ &  $    274 \pm       58$ & $    128 \pm       77$ & $    227 \pm       53$ & $    -18 \pm         37$ & $    0.04 \pm   0.118$  \\  
$  0656+14$ &  $    303 \pm       87$ & $     16 \pm      190$ & $    225 \pm       44$ & $   -101 \pm         26$ & $    0.09 \pm   0.016$  \\  
$  0736-40$ &  $    363 \pm      220$ & $   -250 \pm      265$ & $   -172 \pm       39$ & $    107 \pm         68$ & $    0.15 \pm   0.099$  \\  
$  0740-28$ &  $    274 \pm       72$ & $     22 \pm      183$ & $   -215 \pm       15$ & $     -1 \pm          2$ & $   -0.04 \pm   0.003$  \\  
$  0818-13$ &  $    459 \pm      223$ & $    184 \pm      314$ & $    197 \pm       65$ & $   -284 \pm         66$ & $    0.03 \pm   0.148$  \\  
$  0823+26$ &  $    221 \pm       33$ & $    -20 \pm       50$ & $    119 \pm       21$ & $   -178 \pm         24$ & $   -0.03 \pm   0.127$  \\  
$  0833-45$ &  $    232 \pm      116$ & $     21 \pm      210$ & $   -120 \pm       16$ & $     87 \pm         12$ & $   -0.03 \pm   0.003$  \\  
$  0834+06$ &  $    195 \pm       41$ & $     58 \pm       92$ & $      0 \pm       15$ & $    165 \pm         22$ & $    0.03 \pm   0.112$  \\  
\tableline
\end{tabular}
\end{center}
\end{table}

 \addtocounter{table}{-1}
\begin{table}
\caption{\hfil Pulsar Velocity Components (continued)}
\begin{center}
\begin{tabular}{lrrrrr}
\tableline
\\ 
Name &  $ V\hfil$ &  $ V_r \hfil$ &  $ V_{\perp_1}\hfil$ & $ V_{\perp_2}\hfil$ & $ z_0 \hfil$ \\ 
\cline{2-5}
     & \multicolumn{4}{c}  {$\displaystyle {\rm (km \, s^{-1})} $}  & $(kpc) \hfil$  \\ 
\\ 
\tableline
\tableline 
$  0906-17$ &  $    190 \pm       55$ & $     78 \pm       99$ & $     98 \pm       29$ & $   -108 \pm         31$ & $    0.01 \pm   0.125$  \\  
$  0943+10$ &  $    358 \pm       94$ & $    324 \pm       79$ & $   -114 \pm       62$ & $    -81 \pm         43$ & $    0.04 \pm   0.122$  \\  
$  1133+16$ &  $    466 \pm       90$ & $    -16 \pm       26$ & $   -131 \pm       19$ & $    445 \pm         88$ & $    0.00 \pm   0.123$  \\  
$  1449-64$ &  $    342 \pm      101$ & $     -2 \pm      215$ & $   -175 \pm       17$ & $   -223 \pm         13$ & $   -0.05 \pm   0.019$  \\  
$  1508+55$ &  $    761 \pm       30$ & $    118 \pm       74$ & $   -547 \pm       29$ & $   -510 \pm         26$ & $    0.01 \pm   0.132$  \\  
$  1556-44$ &  $    239 \pm      171$ & $     21 \pm      248$ & $     31 \pm       91$ & $     89 \pm         87$ & $    0.02 \pm   0.146$  \\  
$  1642-03$ &  $    194 \pm       87$ & $    174 \pm       98$ & $     56 \pm       27$ & $    -36 \pm         18$ & $    0.08 \pm   0.123$  \\  
$  1706-16$ &  $   1186 \pm      293$ & $     56 \pm      728$ & $    405 \pm      108$ & $    856 \pm        232$ & $    0.01 \pm   0.125$  \\  
$  1749-28$ &  $    190 \pm      120$ & $     -6 \pm      202$ & $     19 \pm       60$ & $     12 \pm         77$ & $   -0.01 \pm   0.111$  \\  
$  1818-04$ &  $    323 \pm      102$ & $      7 \pm      207$ & $     20 \pm       20$ & $    263 \pm         48$ & $    0.04 \pm   0.079$  \\  
$  1822-09$ &  $    221 \pm      151$ & $    -12 \pm      225$ & $     34 \pm       75$ & $    -91 \pm         75$ & $    0.04 \pm   0.019$  \\  
$  1842+14$ &  $    596 \pm      308$ & $   -348 \pm      401$ & $      9 \pm       44$ & $    400 \pm         75$ & $   -0.11 \pm   0.118$  \\  
$  1911-04$ &  $    240 \pm      114$ & $     21 \pm      211$ & $     98 \pm       60$ & $    -69 \pm         89$ & $   -0.03 \pm   0.171$  \\  
$  1917+00$ &  $    281 \pm      182$ & $     24 \pm      279$ & $    109 \pm       81$ & $    -54 \pm        111$ & $   -0.03 \pm   0.168$  \\  
$  1929+10$ &  $    154 \pm       82$ & $    -30 \pm      154$ & $     67 \pm       10$ & $     33 \pm          5$ & $    0.11 \pm   0.039$  \\  
$  1933+16$ &  $   1010 \pm      309$ & $   -102 \pm      650$ & $     44 \pm       97$ & $   -801 \pm        180$ & $    0.15 \pm   0.086$  \\  
$  1946+35$ &  $    467 \pm      266$ & $     16 \pm      364$ & $   -331 \pm      148$ & $     48 \pm        149$ & $    0.11 \pm   0.174$  \\  
$  1953+50$ &  $   1563 \pm      231$ & $  -1489 \pm      227$ & $   -171 \pm       36$ & $    438 \pm         63$ & $   -0.03 \pm   0.131$  \\  
$  2020+28$ &  $    181 \pm       79$ & $     44 \pm      163$ & $    -52 \pm       17$ & $    -85 \pm         17$ & $   -0.08 \pm   0.047$  \\  
$  2021+51$ &  $    179 \pm       67$ & $    -35 \pm      154$ & $    -56 \pm        2$ & $     92 \pm          4$ & $   -0.04 \pm   0.060$  \\  
\tableline
\end{tabular}
\end{center}
\end{table}

 \addtocounter{table}{-1}
\begin{table}
\caption{\hfil Pulsar Velocity Components (continued)}
\begin{center}
\begin{tabular}{lrrrrr}
\tableline
\\ 
Name &  $ V\hfil$ &  $ V_r \hfil$ &  $ V_{\perp_1}\hfil$ & $ V_{\perp_2}\hfil$ & $ z_0 \hfil$ \\ 
\cline{2-5}
     & \multicolumn{4}{c}  {$\displaystyle {\rm (km \, s^{-1})} $}  & $(kpc) \hfil$  \\ 
\\ 
\tableline
\tableline 
$  2045-16$ &  $    319 \pm       66$ & $   -152 \pm       87$ & $    236 \pm       44$ & $   -127 \pm         43$ & $    0.04 \pm   0.115$  \\  
$  2053+36$ &  $    234 \pm      126$ & $     83 \pm      214$ & $    101 \pm       57$ & $     41 \pm         55$ & $   -0.04 \pm   0.152$  \\  
$  2110+27$ &  $    444 \pm       82$ & $   -257 \pm       96$ & $   -142 \pm       23$ & $   -325 \pm         47$ & $    0.06 \pm   0.116$  \\  
$  2154+40$ &  $   1064 \pm      212$ & $   -942 \pm      206$ & $    483 \pm       77$ & $    -79 \pm         25$ & $    0.04 \pm   0.120$  \\  
$  2217+47$ &  $    403 \pm      160$ & $   -106 \pm      257$ & $   -165 \pm       64$ & $   -276 \pm         51$ & $    0.06 \pm   0.136$  \\  
$  2224+65$ &  $   1647 \pm      224$ & $    166 \pm      560$ & $   1218 \pm      153$ & $    947 \pm        135$ & $    0.03 \pm   0.080$  \\  
$  2351+61$ &  $    392 \pm      126$ & $     24 \pm      236$ & $    323 \pm       53$ & $     74 \pm         25$ & $   -0.04 \pm   0.037$  \\  
\tableline
\end{tabular}
\end{center}
\end{table}

\clearpage

\begin{figure}
\caption{Z-velocity vs. time for objects moving near the Sun
in the Paczynski potential.  Curves are shown only for $V_z > 0$.   
}
\label{fig:vzmodel}
\end{figure}

\begin{figure}
\caption{Lines showing most likely values for $\vr$ and $\zb$.  For three
pulsars (B0540+23, B1449-64 and B2224+65), dashed lines indicate the 68\% confidence interval.
}
\label{fig:zbvr}
\end{figure}

\begin{figure}
\caption{Most likely values for $\vr$ and $\zb$ for four pulsars using three
different values of braking index. Three lines are shown for each value
of braking index to designate the peak likelihood and the 68\% confidence
regions.  
Heavy solid lines:  $n=4$.
Light solid lines:  $n=3$.
Long dashed lines: $n=2.5$.
Dotted lines: $n=2$.
}
\label{fig:zbvr_3psrs}
\end{figure}

\begin{figure}
\caption{Contour plot of the log likelihood function
 for the single Gaussian
model ($n_g.n_h = 1.1$) as a function of the rms velocity and
scale height.  Contours are spaced by $\log 2$.  
}
\label{fig:contour1.1} 
\end{figure}

\begin{figure}
\caption{Contour plots of the log likelihood function
 for the double Gaussian
model ($n_g.n_h = 2.1$) as a function of values
for pairs of parameters and for 2D slices through the 
best-fit model.  Contours are spaced by $\log 2$. 
}
\label{fig:contour2.1} 
\end{figure}

\begin{figure}
\caption{Marginalized probability density functions
for the four parameters of the
double Gaussian model, $n_g.n_h = 2.1$.
}
\label{fig:marg}
\end{figure}

\begin{figure}
\caption{
One-dimensional cuts through the likelihood functions for
a double-Gaussian velocity pdf where the torque decay time $\tau_K$
is allowed to vary for different values of braking index.
The heavy solid line for n = 2.5 yields the maximum likelihood
solution at $\tau_K \sim 3$ Myr.  The two heavy dashed lines,
for n = 4 and 4.5, yield the maximum likelihood if there is
no decay, $\tau_K\to\infty$. 
}
\label{fig:ntau}
\end{figure}

\begin{figure}
\caption{
(Top) Cumulative distribution functions for the velocity magnitude
at birth using best fit parameter values for models 1.1 (dotted line), 
2.1 (heavy solid line), and 3.1 (dashed line).  
The light solid lines represent model 2.1
evaluated with parameter values that are $\pm 1\sigma$ from the
best fit values.  The vertical dashed line marks 500 km s$^{-1}$,
the nominal speed of escape from the Galaxy at the solar circle.  
(Bottom) Differential probability density functions for models
1.1, 2.1 and 3.1 using best fit parameter values .
}
\label{fig:pdf_cdf}
\end{figure}

\begin{figure}
\caption{
Scatter plots of radial velocity $\vr$ against birth altitude, $\zb$,
and perpendicular speed, $\vperp$, for three pulsars.  The 
values result from the Monte Carlo solutions derived using 
Eq.~\ref{eq:pdf_zv}.
(a) B1953+50;
(b) B2154+40;
(c) B2224+65
}
\label{fig:scatterplots}
\end{figure}

\begin{figure}
\caption{Probability density functions and cumulative distribution
functions for the ages of three pulsars.   Vertical lines indicate 
the conventional spindown ages obtained from Eq.~\ref{eq:tau}
with a braking index, $n=3$.
}
\label{fig:agepdfcdf}
\end{figure}
 
\begin{figure}
\caption{Plot showing the acceptable range of possible ages for
each of 47 pulsars, calculated using the pdf of Eq.~\ref{eq:pdft}.
The horizontal line indicates the interval enclosing 68\% of 
the probability.  The $\times$ indicates the median of each
pdf, the open circle the mode of the pdf, and  
the solid circle indicates the spindown
age $\taus \equiv P /2\dot P$. For the Crab and Vela pulsars,
the kinematic age is unconstraining because of their
young ages; for them, we show only their spindown times.
}
\label{fig:ages}
\end{figure}

\begin{figure}
\caption{
Probability density for the ratio of pulsar age to spindown
age, $r \equiv t /\taus$, based on the individual pdfs
for 45 pulsars.  The Crab and Vela pulsars have been excluded
from our sample because their kinematic ages are poorly constrained. 
}
\label{fig:sumpdfages}
\end{figure}

\begin{figure}
\caption{Plot of correlation coefficient between log V and
log $P^{\alpha}\dot P^{\beta}$ vs. $\beta$. 
}
\label{fig:corr}
\end{figure}
 
\vfill


\begin{thebibliography}{}

\def\nature{{Nature}}
\def\apjsuppl{{ApJSuppl}}

\bibitem[Anderson \& Lyne 1983]{al83}
Anderson, B. \&  Lyne, A.G. 1983, \nature, 303, 597. 

\bibitem[Arzoumanian \etal 1996]{aptw96}
Arzoumanian, Z., Phillips, J.A., Taylor, J.H. \& Wolszczan, A.
1996, ApJ, 470, 1111.

\bibitem[Bailes  1989]{b89}
Bailes, M.  1989, \apj, 342, 917                                    

\bibitem[Bailes \etal 1990]{b+90}
Bailes, M. \etal 1990, \mnras, 247, 322                                    

\bibitem[Bailes \etal 1989]{b+89a}
Bailes, M., Reynolds, J.E., Manchester, R. N., Kesteven, M.J., \& Norris, R.P. 
 1989, \apjl, 343, 53L.

\bibitem[Bailes \etal 1989]{b+89b}
Bailes, M., Manchester, R. N., Kesteven, M.J., Norris, R.P. \& Reynolds, J.E.
 1989, \apj, 342, 917.

\bibitem[Bethe 1993]{b93}
Bethe, H. A. 1993, ApJ, 412, 192.
 
\bibitem[Bhattacharya \& van den Heuvel 1991]{bv91}
Bhattacharya, D. \& van den Heuvel, E. P. J. 1991, Phys. Rep. 203, 1.

\bibitem[Binney \& Tremaine 1987]{bt87}
Binney, J. \& Tremaine, S. 1987, {\it Galactic Dynamics},
(Princeton: Princeton University Press), p. 17. 

\bibitem[Birkel \& Toldra]{bt97}
Birkel, M. \& Toldr\`{a}, R. 1997, astro-ph/9704138

\bibitem[Bisnovatyi-Kogan 1996]{bk96}
Bisnovatyi-Kogan, G. S. 1996, in 
in High Velocity Neutron Stars \& Gamma-Ray Bursts,
eds. R. E. Rothschild \& R. E. Lingenfelter, 38.

\bibitem[Blandford, Applegate \& Hernquist]{bah83}
Blandford, R. D., Applegate, J. H. \& Hernquist, L. 1983, \mnras, 204, 1025.

\bibitem[Blauuw 1961]{b61}
Blaauw, A. 1961 B.A.N. 15, 265.

\bibitem[Boyd \etal 1995]{b+95}
Boyd, P.T. \etal 1995, \apj, 448, 365

\bibitem[Burderi \etal 1996]{bkw96}
Burderi, L., King, A.R., \& Wynn, G.A. 1996,
\mnras, 283, L63

\bibitem[Burrows, Hayes \& Fryxell 1995]{bhf95}
Burrows, A., Hayes, J. \& Fryxell, B. A. 1995,
\apj, 450, 330.

\bibitem[Burrows \& Hayes  1996]{bh96}
Burrows, A. \& Hayes, J 1996,
PhysRevLett, 76, 352.

\bibitem[Camilo \etal 1996]{c+96}
Camilo, F., Nice, D.J. \& Taylor, J.H. \etal 1996, \apj, 461, 812.

\bibitem[Camilo \etal 1994]{c+94}
Camilo, F., Thorsett, S. \& Kulkarni, S.R. 1994, \apjlett, 421, L15.

\bibitem[Campbell et al. 1996]{cam+96}
Campbell, R.M.  \etal 1996, \apjl, 461, 95.   

\bibitem[Caraveo 1993]{c93}
Caraveo, P.A.  1993, \apjl, 415, L111

\bibitem[Caraveo et al. 1996]{cb+96}
Caraveo, P.A.  \& Bignami, G.F., Mignami, R., Taff, L.G. 1996, \apjl, 461, 91. 

\bibitem[Chernoff, Cordes \& Arzoumanian 1996]{cca96}
Chernoff, D., Cordes, J. M. and Arzoumanian, Z. 1996,
in High Velocity Neutron Stars \& Gamma-Ray Bursts,
eds. R. E. Rothschild \& R. E. Lingenfelter, 16.

\bibitem[Cordes 1986]{c86}
Cordes, J. M. 1986, \apj, 311, 183.  
                                       
\bibitem[Cordes 1987]{c87}
Cordes, J. M. 1987, in The Origin \& Evolution of Neutron Stars,
Proceedings of the 125th Symposium of the IAU held
 in Nanjing, China, eds. D.J. Helfand \& J. H. Huang, 35.  

\bibitem[Cordes, Romani \& Lundgren 1993]{crl93}
Cordes, J.M., Romani, R. W. \& Lundgren, S. C. 1993, \nature, 362, 133.

\bibitem[Cordes \& Chernoff 1997]{cc97}
Cordes, J.M. \& Chernoff, D. 1997, \apj, 482, 971 (Paper I).

\bibitem[Cordes \& Wasserman 1984]{cw84}
Cordes, J. M. \& Wasserman, I. 1984, \apj, 279, 798.

\bibitem[Cordes, Wasserman \& Blaskiewicz 1990]{cwb90}
Cordes, J. M., Wasserman, I \& Blaskiewicz, M. 1990, \apj, 349, 546.

\bibitem[Cunha \& Smith 1996]{cs96}
Cunha, K. \& Smith, V. V. 1996, \aap, 309, 892.

\bibitem[Dewey \& Cordes 1987]{dc87}
Dewey, R.~J. \& Cordes, J.~M. 1987, { \apj},  { 321},  780.

\bibitem[Downs \& Reichley 1983]{dr83}
Downs, G. S. \& Reichley P. E. 1983, \apjsuppl, 53, 169.

\bibitem[Fomalont et al. 1992]{f+92}
Fomalont, E. \etal  1992, \mnras, 258, 497.  

\bibitem[Frail 1990]{f90}
Frail, D. A. 1990, PhD Thesis, University of Toronto.

\bibitem[Frail \& Weisberg 1990]{fw90}
Frail, D.A. \& Weisberg, J. 1990, \aj, 100, 743.


\bibitem[Frail, Goss \& Whiteoak 1994]{fgw94}
Frail, D. A., Goss, W. M. \& Whiteoak, J. B. Z. 
1994, \apj, 437, 781.

\bibitem[Frisch 1993]{f93}
Frisch, P. C. 1993, \nature, 364, 395.

\bibitem[Gehrels \& Chen 1003]{gc93}
Gehrels, N. \& Chen, W. 1993, \nature, 361, 706.

\bibitem[Goldreich \& Julian 1970]{gj70}
Goldreich, P. \& Julian, W. H. 1970, \apj, 160, 971.

\bibitem[Goldreich, Lai \& Sahrling 1997]{gls97}
Goldreich, P., Lai, D. \& Sahrling, M. 1997, preprint.

\bibitem[Goldreich \& Reisenegger 1992]{gr92}
Goldreich, P. \& Reisennegger, A. 1992, \apj, 395, 250.


\bibitem[Gott, Gunn \& Ostriker 1970]{ggo70}
Gott, J. R., Gunn, J. E. \& Ostriker, J. P. 1970,
\apjlett, 160, L91.

\bibitem[Gunn \& Ostriker 1970]{go70}
Gunn, J. E. \& Ostriker, J. P. 1970, \apj, 160, 979.

\bibitem[Gregory \& Loredo 1992]{gl92}
Gregory, P.~C. \& Loredo, T.~J. 1992, { \apj}, { 398}, 146.

\bibitem[Gwinn et al. 1986]{g+96}
Gwinn, C.R., Taylor, J. H., Weisberg, J.M. \& Rawley, L.A.  1986,
 \aj, 91, 974.

\bibitem[Halpern \& Holt 1992]{hh92}
Halpern, J. P. \& Holt, S. S. 1992, \nature, 357, 222.

\bibitem[Hammersky \etal 1995]{hgmc95}
Hammersky, P.L., Garzon, F., Mahoney, T., Calbet, X. 1995,
\mnras, 273, 206.

\bibitem[Harrison \& Tademaru 1975]{ht75}
Harrison, E. R. \& Tademaru, E. 1975, \apj, 201, 447.

\bibitem[Harrison \etal 1993]{hla93}                                        
Harrison, P.A., Lyne, A.G. \& Anderson, B. 1993,
\mnras, 261, 113

\bibitem[Heintzmann \& Schruefer 1982]{hs82}
Heintzmann, H. \& Schruefer, E. 1982, \aap, 111, L4

\bibitem[Helfand \& Tademaru 1977]{ht77}
Helfand, D. J. \& Tademaru, E. 1977, \apj, 216, 842.

\bibitem[Herant, Benz \& Colgate 1992]{hbc92}
Herant, M., Benz, W. \& Colgate, S. 1992, \apj, 395, 642.

\bibitem[Horowitz \& Piekarewicz 1997]{hp97}
Horowitz, C. J. \& Piekarewicz, J. 1997, hep-ph/9701214.

\bibitem[Hulse and Taylor 1975]{hut75}
Hulse, R.~A. and Taylor, J.~H. 1975, { \apjlett}, { 195}, L51.

\bibitem[Iben \& Tutukov 1996]{it96}
Iben, Jr., I. \& Tutukov, A. V. 1996, \apj, 456, 738.

\bibitem[Itoh \etal 1995]{ikh95}
Itoh, N., Kotouda, T. \& Hiraki, K. 1995, \apj, 455, 244.

\bibitem[Janka \& M\"uller 1994]{jm94}
Janka, H. Th. \& M\"uller, E. 1994, \aap, 290, 496.

\bibitem[Kaspi \etal 1995]{k+95}
Kaspi, V., Manchester, R. N., Siegman, B., Johnston, S.
\& Lyne, A.G. 1994, \apjl, 422, L83 

\bibitem[Kaspi \etal 1996]{k+96}
Kaspi, V., Bailes, M., Manchester, R.N., Stappers, B.W.
\& Bell, J.F. 1996, \nature, 381, 584

\bibitem[Koribalski \etal 1995]{kor+95}
Koribalski, B., Johnston, S., Weisberg, J. M. \& Wilson, W.  1995, 
\apj, 441, 756

\bibitem[Kulkarni 1986]{k86}
Kulkarni, S. R. 1986, \apjl, 306, L85 

\bibitem[Kusenko \& Segre 1996]{ks96}
Kusenko, A. \& Segr\`{e} 1996a, Phys. Rev. Lett. 77, 4872.

\bibitem[Kusenko \& Segre 1996]{ks96+}
Kusenko, A. \& Segr\`{e} 1996b, astro-ph/9608103

\bibitem[Lai, Bildsten \& Kaspi 1995]{lbk95}
Lai, D., Bildsten, L. \& Kaspi, V. M. 1995, \apj, 452, 819

\bibitem[Leonard \& Tremaine 1990]{lt90}
Leonard, P.J.T. \& Tremaine, S. 1990, { \apj}, { 353}, 486.

\bibitem[Lorimer, Lyne \& Anderson 1995]{lla95}
Lorimer, D. R., Lyne, A.G. \& Anderson, B. 1995, \mnras

\bibitem[Lyne, Anderson \& Salter 1982]{las82}
Lyne, A.G., Anderson, B. \& Salter, M. J. 1982,
\mnras, 213, 613.                             

\bibitem[Lyne \& Lorimer 1994]{ll94}
Lyne, A.G. \& Lorimer, D. R. 1994, \nature, 369, 127.

\bibitem[Lyne \& Manchester 1988]{lm88}
Lyne, A.G. \& Mancheter, R. N. 1988, \mnras, 234, 477

\bibitem[Lyne et al. 1988]{lps88}
Lyne, A.G., Pritchard, R. S. \& Graham-Smith, F.  1988,
\mnras, 233, 667


\bibitem[Lyne et al. 1996]{lpgc96}
Lyne, A.G., Pritchard, R. S., Graham-Smith, F. \& Camilo, F. 1996,
\nature, 381, 497

\bibitem[Lyne, Manchester \& Taylor 1985]{lmt85}
Lyne, A.G., Manchester, R. N. \& Taylor, J.H. 1985, 
\mnras, 213, 613.

\bibitem[Madau \& Blaes 1994]{mb94}
Madau, P. \& Blaes, O. 1994, \apj, 423, 748.

\bibitem[Manchester, Newton \& Durdin 1985]{m+85}
Manchester, R.N., Newton, L. M. \& Durdin, J.M. 1985,
\nature, 313, 374  

\bibitem[Manning, Jeffries \& Willmore 1996]{mjw96}
Manning, R. A., Jeffries, R. D. \& Willmore, A. P. 1996, \mnras, 278, 577.

\bibitem[Melatos 1997]{m97}
Melatos, A. 1997, preprint.

\bibitem[Michel 1969]{m69}
Michel, F. C. 1969, \apj, 158, 727.

\bibitem[Mihalas \& Binney 1981]{mb81}
Mihalas, D. \& Binney, J. 1981,
{\it Galactic Astronomy, Structure and Kinematics},
W. H. Freeman, New York, pp. 382-383.

\bibitem[Mollerach \& Roulet 1997]{mr97}
Mollerach, S. \& Roulet, E. 1997, \apj, 479, 147.

\bibitem[Muslimov \& Page]{mp96}
Muslimov, A. \& Page, D. 1996, \apj, 458, 347.

\bibitem[Narayan \& Ostriker 1990]{no90}
Narayan, R. \& Ostriker, J. P. 1990, \apj, 352, 222.

\bibitem[Ostriker \& Gunn 1969]{og69}
Ostriker, J. P. \& Gunn, J. E. 1969, \apj, 157, 1395.

\bibitem[Paczynski 1990]{p90}
Paczynski, B.  1990 \apj, { 348}, 485.

\bibitem[Portegies Zwart \& Verbunt]{pzv96}
Portegies Zwart, S.F. \& Verbunt, F. 1996, \aap, 309, 179P. 

\bibitem[Prentice \& ter Haar 1969]{ptH69}
Prentice, A. \& ter Haar, D. 1969, \mnras, 146, 423

\bibitem[Press \etal 1993]{p+93}
Press, W.H., Flannery, B. P., Teukolsky, S. A. \& Vetterling, W. T. 1992,
{\it Numerical Recipes in FORTRAN}, 2nd edition, (New York: Cambridge
University Press).

\bibitem[Qian 1997]{q97}
Qian, Y.-Z. 1997, astro-ph/9705055.

\bibitem[Radhakrishnan \& Shukre 1985]{rs85}
Radhakrishnan, V. \& Shukre, C. S. 1985, 
in Supernovae: Their Progenitors and Remnants,
ed. G. Srinivasan \& V. Radhakrishnan (Bangalor: Indian
Acad. Sci), 155. 

\bibitem[Romani 1990 ]{r90}
Romani, R. W. 1990, \nature, 347, 741 

\bibitem[Rubinstein 1981]{r81}
Rubinstein, R. 1981, {\it Simulation and the Monte Carlo Method},
John Wiley \& Sons, New York.

\bibitem[Ruderman 1991]{r91}
Ruderman, M. 1991, \apj, 366, 261.

\bibitem[Sang \& Chanmugam 1990]{sc90}
Sang, Y. \& Chanmugam, G. 1990, \apj, 363, 597.

\bibitem[Shklovskii 1970]{s70}
Shklovskii, I. S. 1970, Astr. Zh., 46, 715.

\bibitem[Smith, Cunha \& Plez 1994]{scp94}
Smith, V.V., Cunha, K. \& Plez, B. 1994, \aap, 281, L41

\bibitem[Taam \& van den Heuvel 1986]{tvdh86}
Taam, R.E. \& van den Heuvel, E.P.J. 1986, \apj, 305, 235

\bibitem[Taylor \& Cordes 1993]{tc93}
Taylor, J.~H. \& Cordes, J.~M. 1993, { \apj},
  { 411}, 674.


\bibitem[Taylor, Manchester, \& Lyne 1993]{tml93}
Taylor, J.~H., Manchester, R.~M., \& Lyne, A.~G. 1993, { \apjsupp}, 
  { 88}, 529.  


\bibitem[Thompson \& Cordova 1994]{tc94}
Thompson \& Cordova, 1994, \apjlett, 421, L13 

\bibitem[Urpin \& Geppert 1995]{ug95}
Urpin, V. \& Geppert, U. 1995, \mnras, 275, 1117

\bibitem[van den Heuvel 1993]{vdh93}
van den Heuvel, E.P.J. 1993, 
in Planets around pulsars: Proceedings of the Conference,
California Inst. of Technology, 123.

\bibitem[Vivekanand \& Narayan 1981]{vn81}
Vivekanand, M. \& Narayan, R. 1981, 
Journal of Astrophysics and Astronomy,
2, 315

\bibitem[Wang 1996]{wang96}
Wang, J.C.L. 1996, \nature, 379, 206.

\bibitem[Wang 1997]{wang97}
Wang, J.C.L. 1997, preprint.

\bibitem[Wendell 1997]{w97}
Wendell, C.E. 1997, \apjlett, 333, L95.

\bibitem[Weisberg, Romani \& Taylor 1989]{wrt89}
Weisberg, J. M., Romani, R. W. \& Taylor, J. H. 1989, \apj, 347, 1030.

\bibitem[Young \& Chanmugam 1995]{yc95}
Young, E.J. \& Chanmugam, G. 1995, \apjlett, 442, L53
\end{thebibliography}
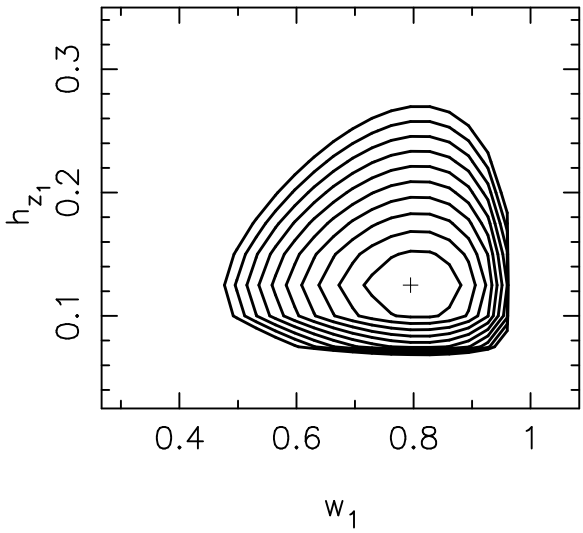
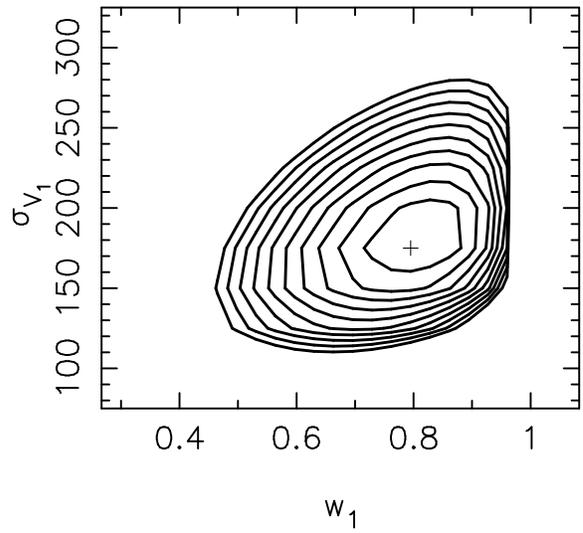
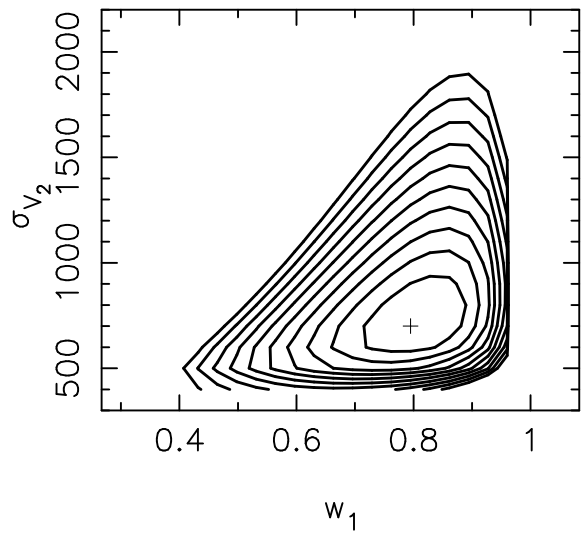
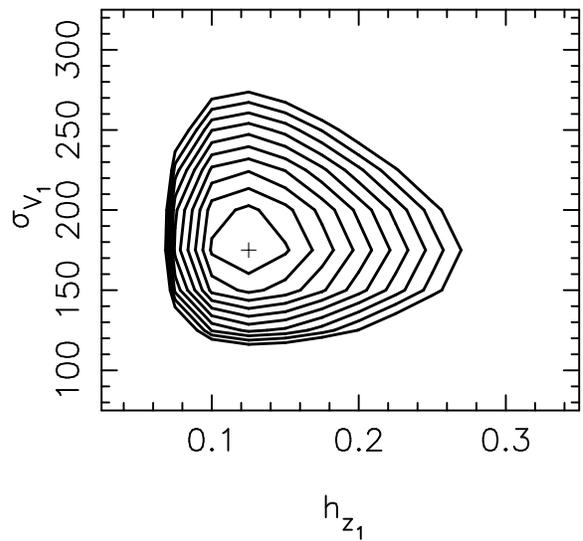
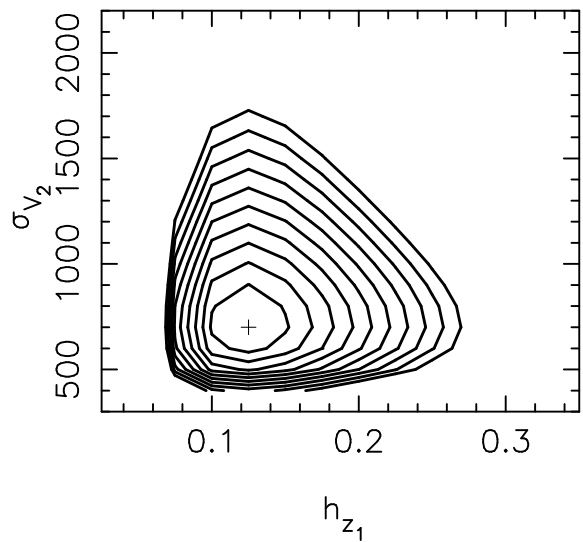
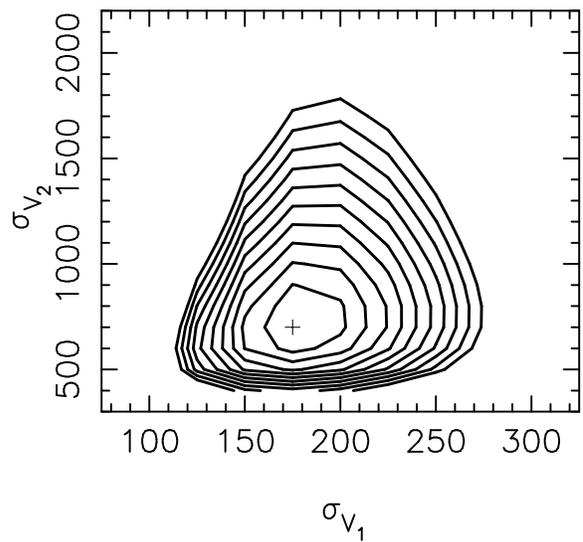
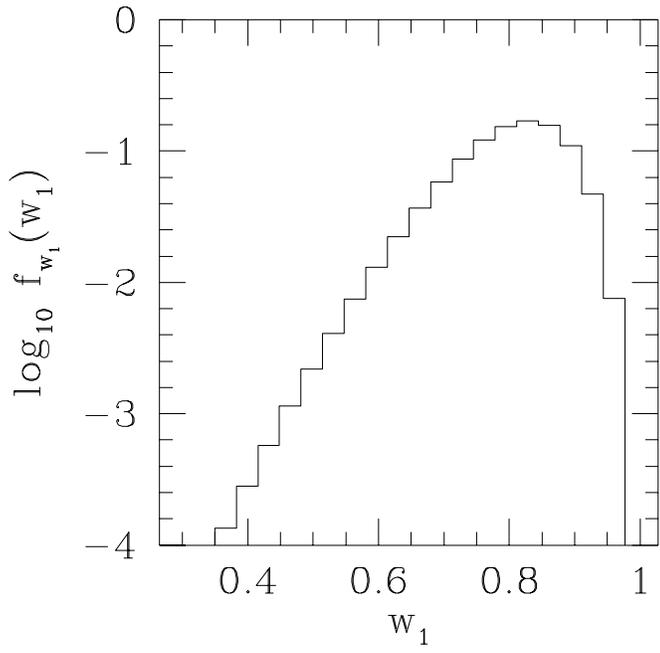
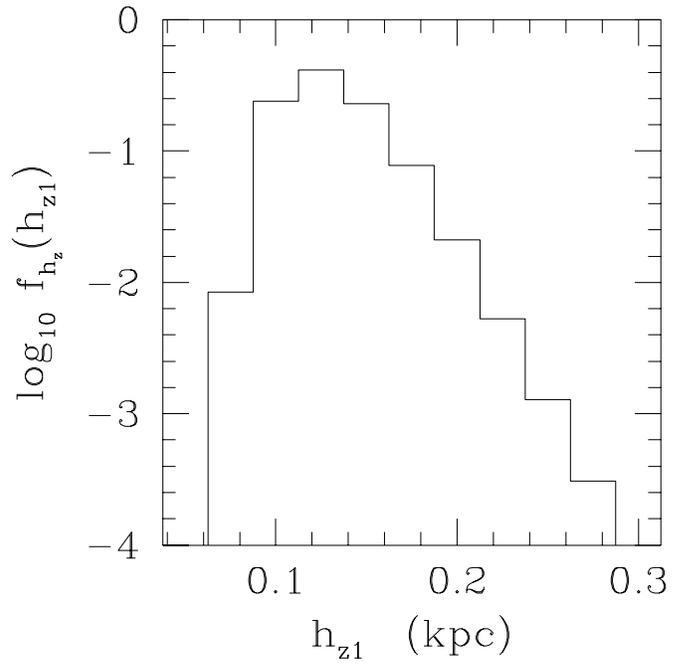
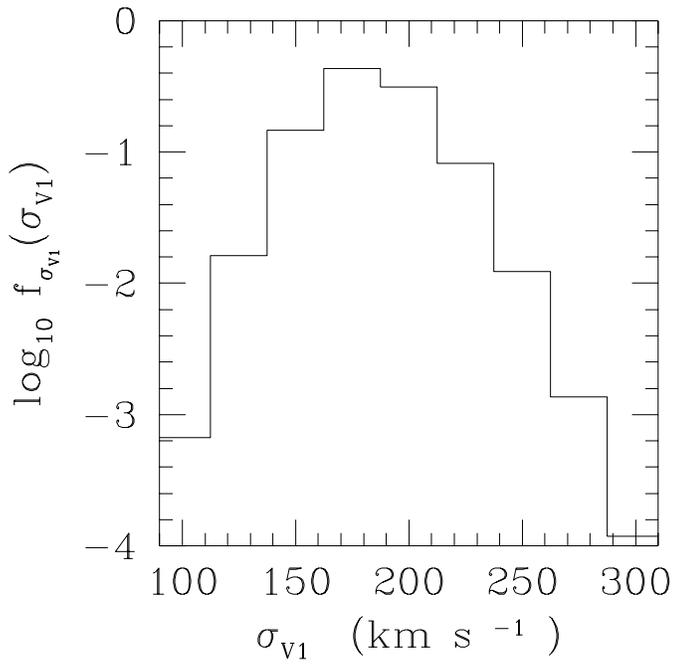
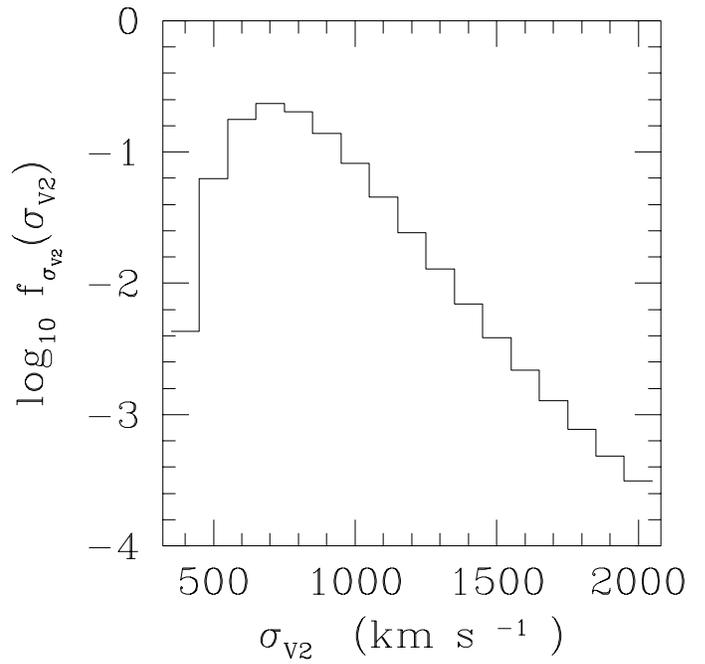
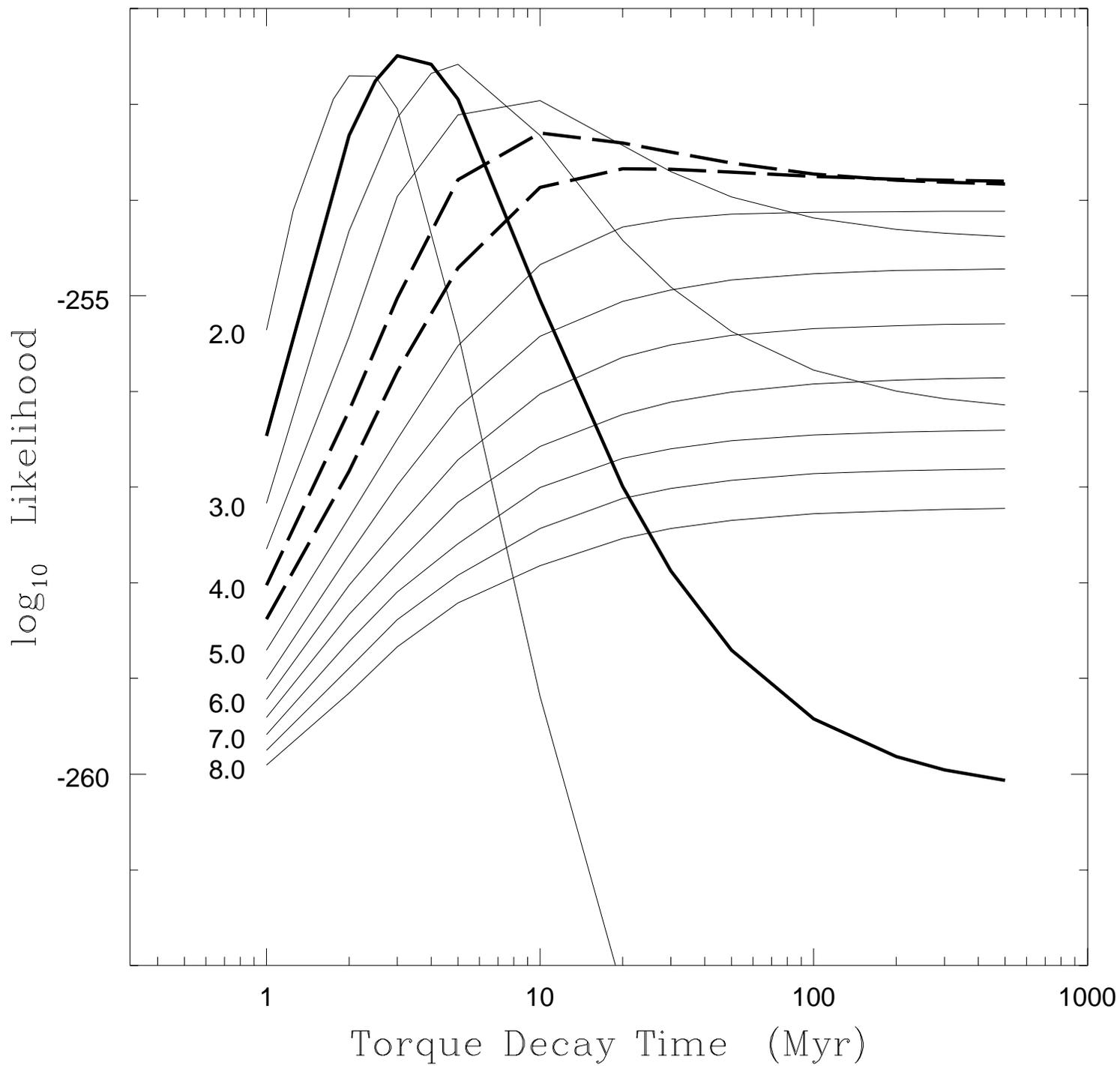
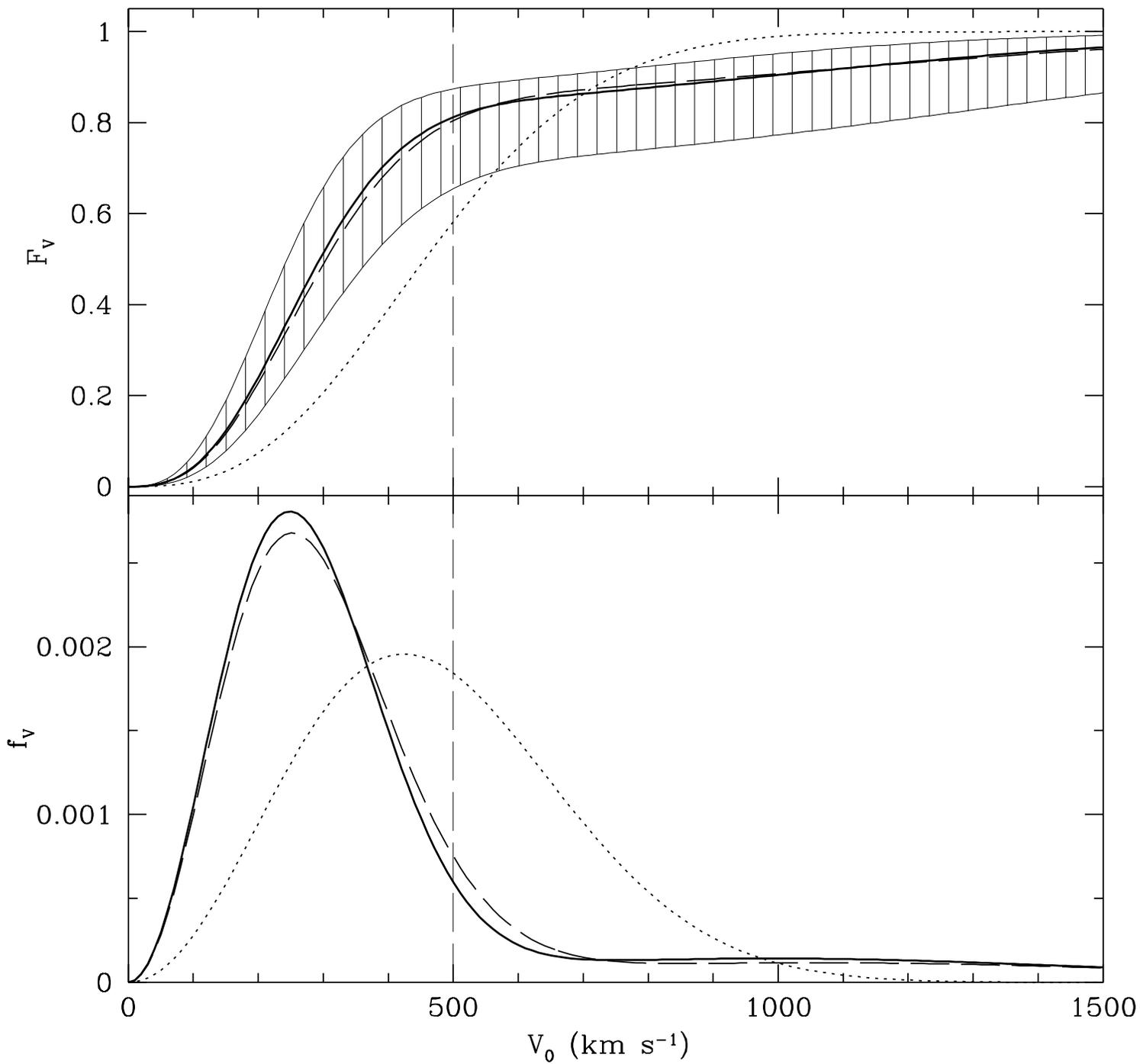
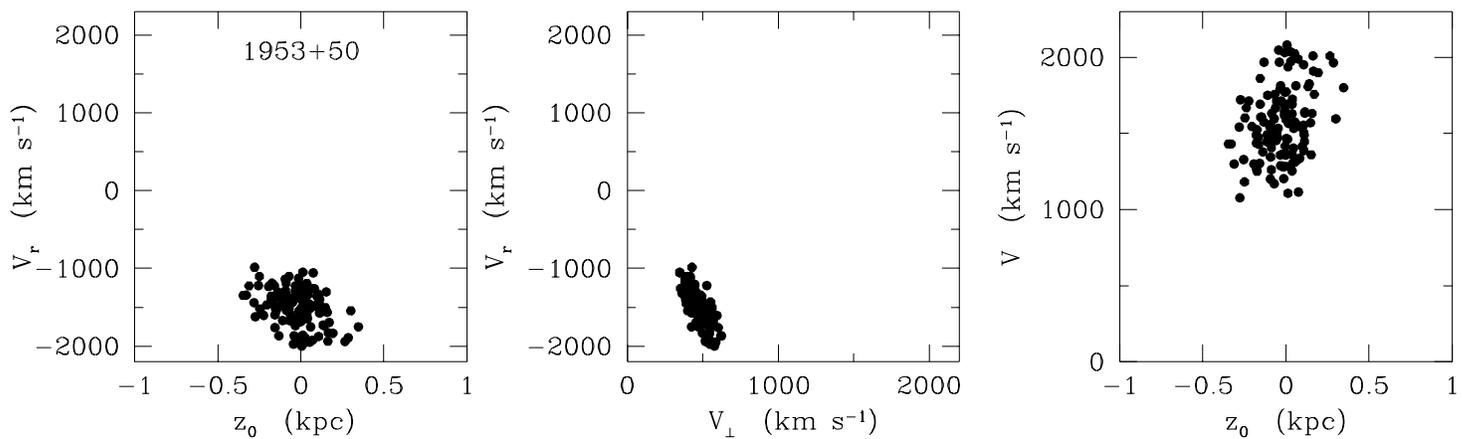
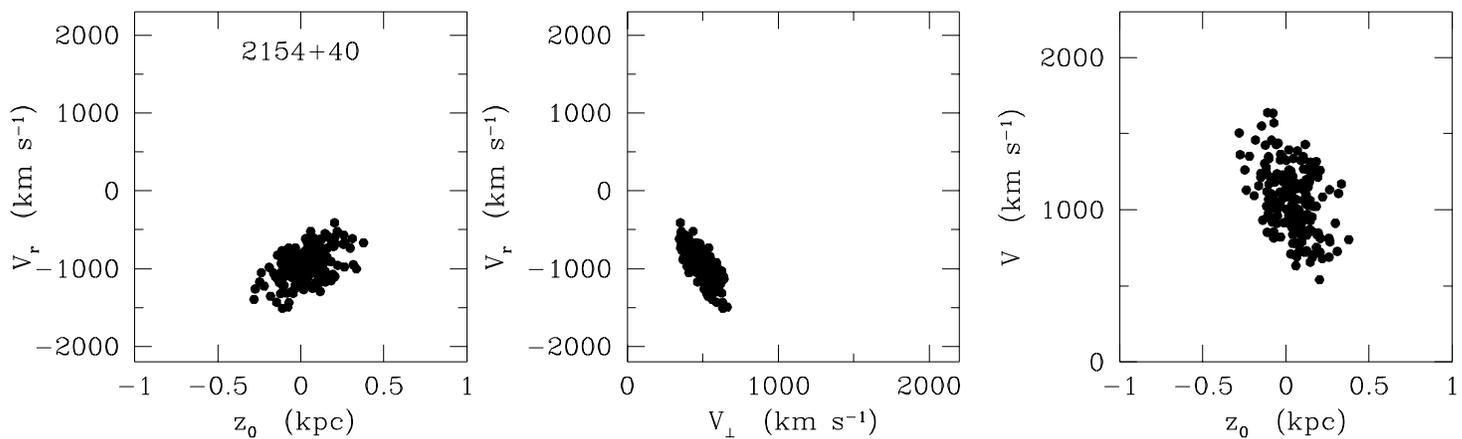
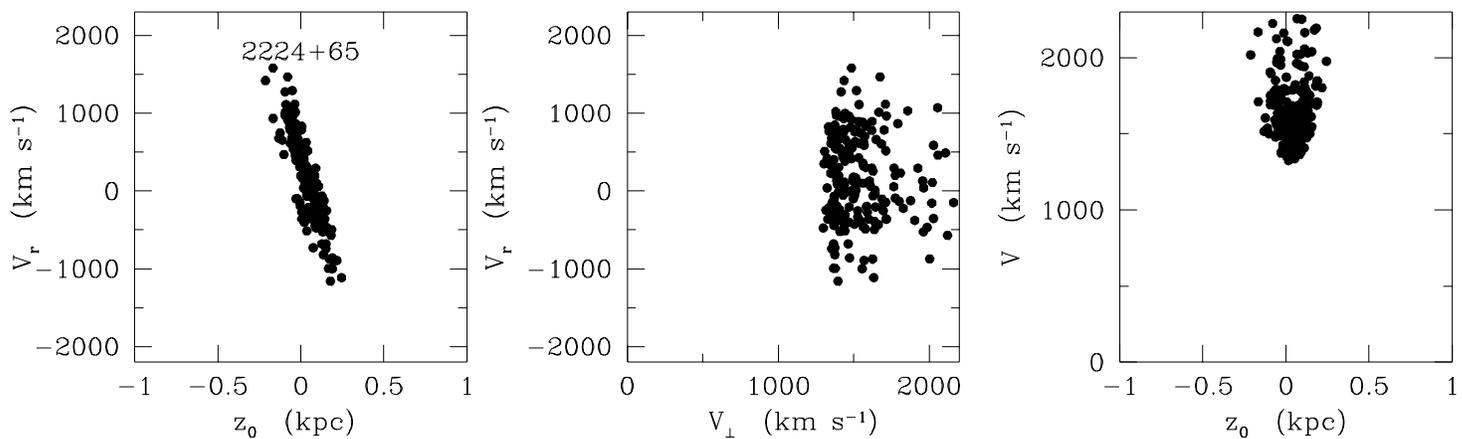
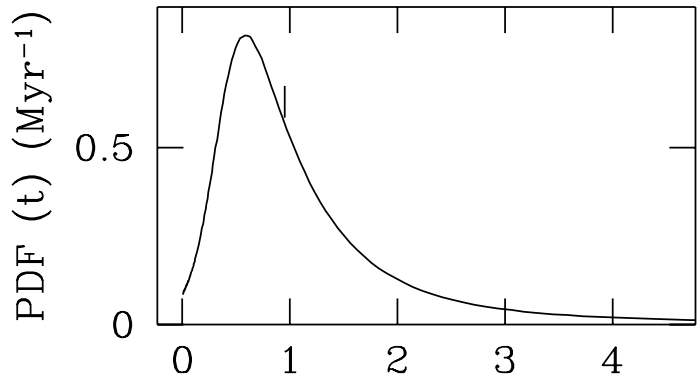
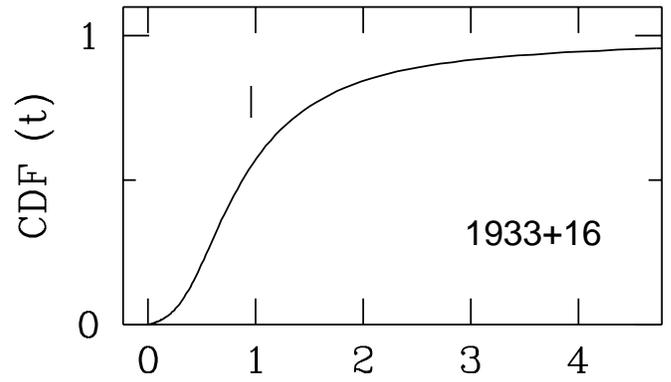
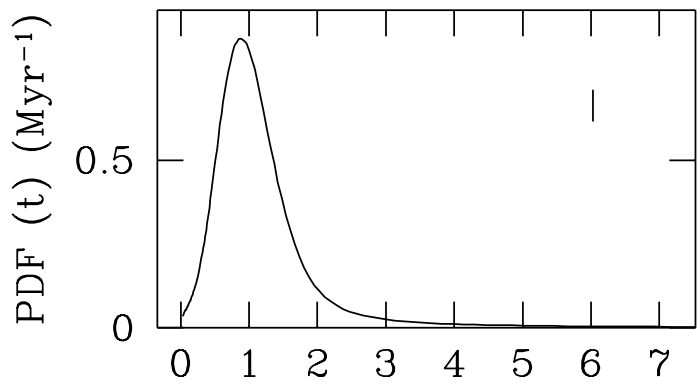
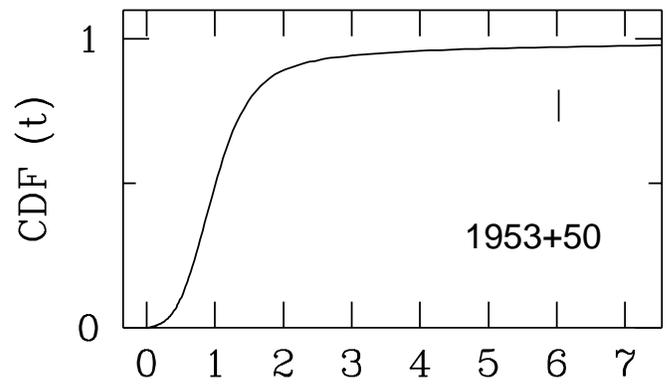
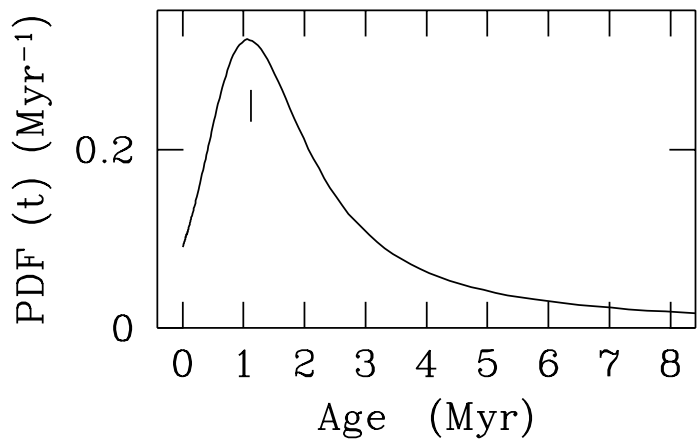
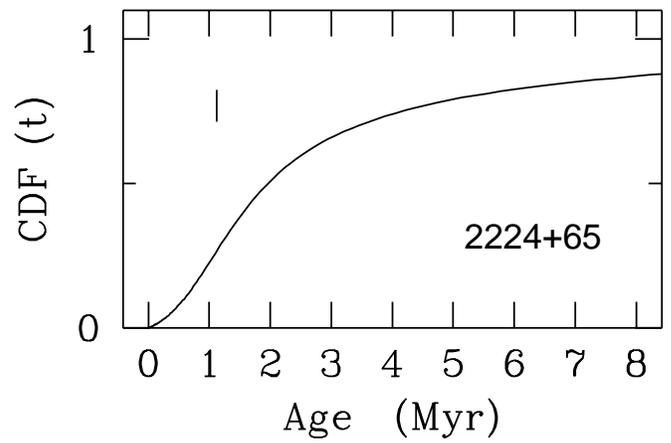
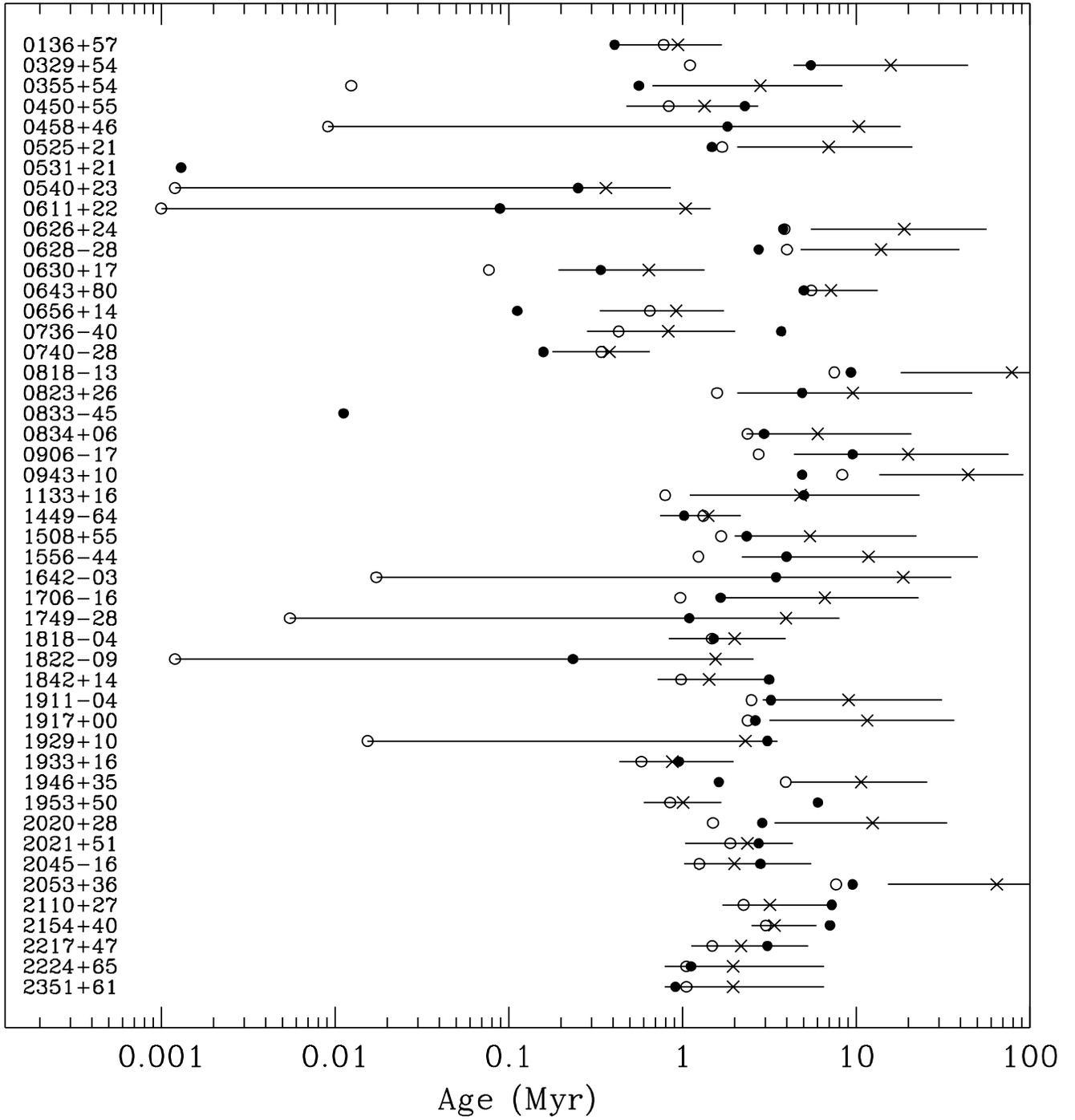
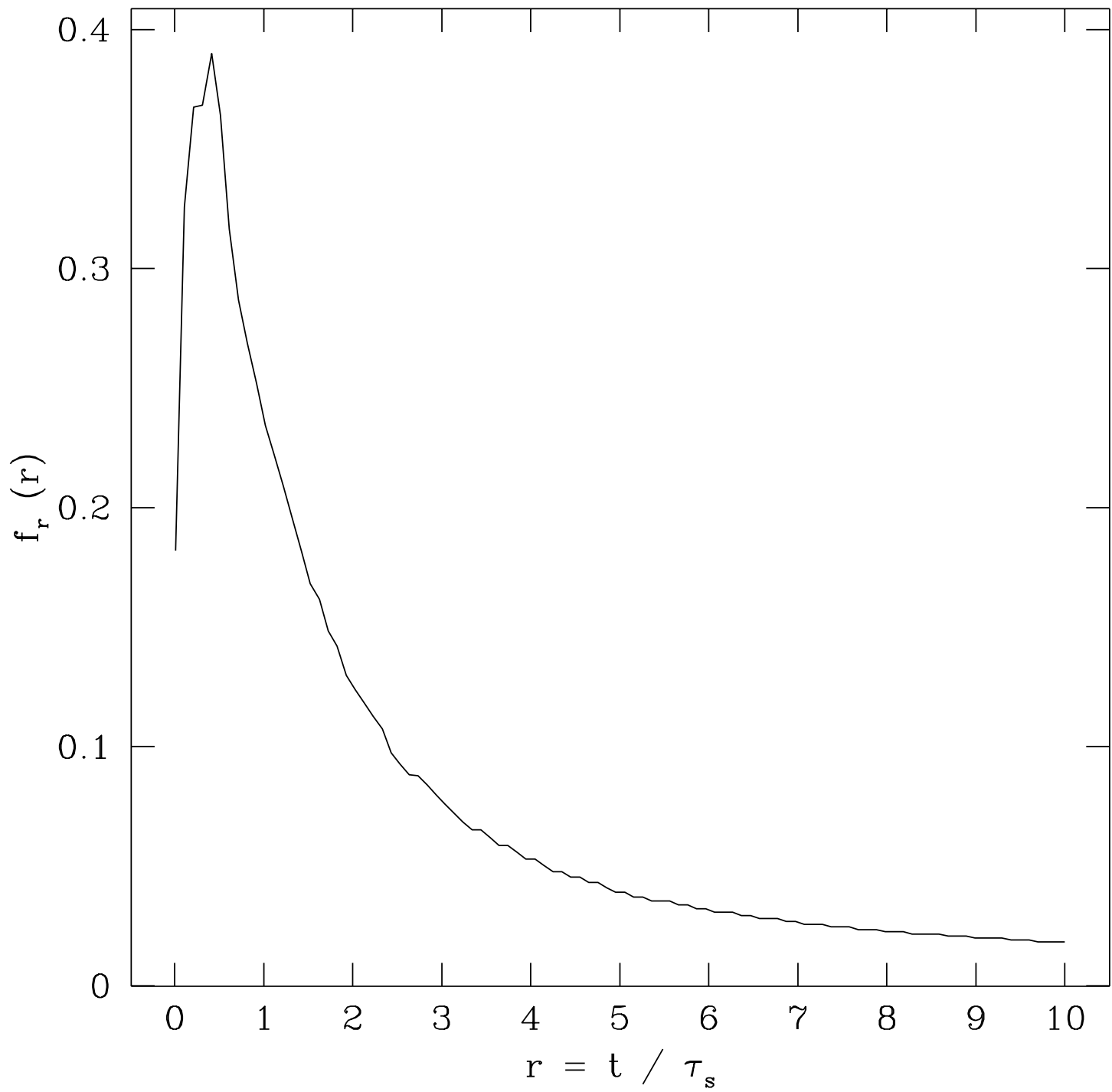
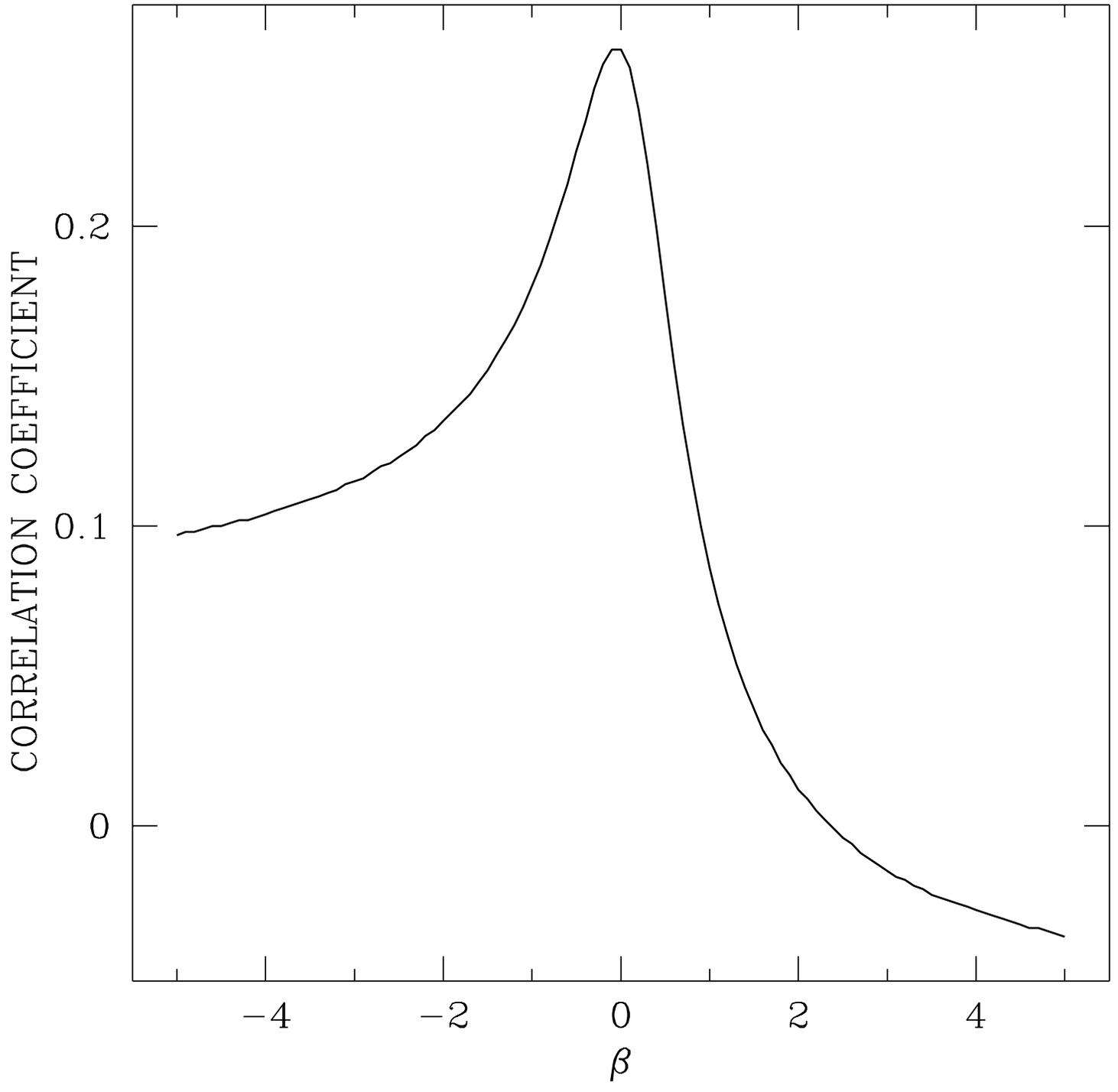
\end{document}